\documentclass[reprint, aps, prl, nofootinbib, superscriptaddress, fleqn]{revtex4-2}
\usepackage[utf8]{inputenc}
\usepackage{amsmath}
\usepackage{amssymb}
\usepackage{graphicx}
\usepackage[dvipsnames]{xcolor}
\usepackage{calc}
\usepackage{etoolbox}
\usepackage{booktabs}
\usepackage{tabularx}
\usepackage{multirow}
\usepackage[normalem]{ulem}
\usepackage{siunitx}
\usepackage{mleftright}
\usepackage{hyperref}
\usepackage[capitalise]{cleveref}

\makeatletter
\input{aas_macros.sty}
\let\jnl@style=\relax
\makeatother

\makeatletter
\AtBeginDocument{\immediate\write\@auxout{\string\citation{apsrev42Control}}}
\makeatother

\newlength{\affilindent}
\newcommand*{\affilscriptphantom}{\phantom{\normalfont\textsuperscript{99}}}
\newcommand*{\printaffiliation}[4]{%
    \ifnumequal{#1}{0}{}{%
        \setlength{\affilindent}{\widthof{\affilscriptphantom}}%
        \let\\=\newline%
        \par\noindent%
        \looseness=-100\hangindent=\affilindent%
        \affilscriptphantom\llap{\normalfont\textsuperscript{#1}}%
        \ignorespaces#3%
        \par%
    }%
}

\makeatletter
\appto{\frontmatter@above@affiliation@script}{%
    \global\let\affiliationslist=\@AFF@list%
    \let\@AFF@list=\relax%
    Affiliations are given after the Appendix.%
}{}{}
\newcommand{\printaffiliations}{%
    \begingroup%
    \let\AFF@opr=\printaffiliation%
    \frontmatter@affiliationfont%
    \raggedright%
    \affiliationslist%
    \endgroup%
}
\makeatother

\DeclareSIUnit{\year}{yr}
\DeclareSIUnit{\Gyr}{\giga\yr}
\DeclareSIUnit{\pc}{pc}
\DeclareSIUnit{\kpc}{\kilo\pc}
\DeclareSIUnit{\Mpc}{\mega\pc}
\DeclareSIUnit{\Gpc}{\giga\pc}
\DeclareSIUnit{\hHubble}{\text{\ensuremath{h}}}
\DeclareSIUnit{\Msun}{\text{\ensuremath{M_\odot}}}
\newcommand*{\eqlabelleft}{(}
\newcommand*{\eqlabelright}{)}
\NewDocumentCommand{\pcref}{m}{%
    \begingroup%
    \renewcommand*{\eqlabelleft}{}%
    \renewcommand*{\eqlabelright}{}%
    \cref{#1}%
    \endgroup%
}
\creflabelformat{equation}{\eqlabelleft#2#1#3\eqlabelright}

\begin{document}

\title{Measurement of Parity-Violating Modes of the  Dark Energy Spectroscopic Instrument (DESI) Year 1 Luminous Red Galaxies' 4-Point Correlation Function}

\author{Zachary Slepian}
\affiliation{Department of Astronomy, University of Florida,\\211~Bryant Space Science Center, Gainesville, FL 32611, USA}

\author{Alex Krolewski}
\affiliation{Department of Physics and Astronomy, University of Waterloo,\\200~University Ave.\ W, Waterloo, ON N2L 3G1, Canada}
\affiliation{Waterloo Centre for Astrophysics, University of Waterloo,\\200~University Ave.\ W, Waterloo, ON N2L 3G1, Canada}

\author{Alessandro Greco}
\affiliation{Department of Astronomy, University of Florida,\\211~Bryant Space Science Center, Gainesville, FL 32611, USA}

\author{Simon May}
\affiliation{Perimeter Institute for Theoretical Physics,\\31~Caroline St.\ North, Waterloo, ON N2L 2Y5, Canada}

\author{William Ortol\'{a} Leonard}
\affiliation{Department of Physics, University of Florida,\\2001 Museum Road, Gainesville, FL 32611, USA}

\author{Farshad Kamalinejad}
\affiliation{Department of Physics, University of Florida,\\2001 Museum Road, Gainesville, FL 32611, USA}

\author{Jessica Chellino}
\affiliation{Department of Astronomy, University of Florida,\\211~Bryant Space Science Center, Gainesville, FL 32611, USA}

\author{Matthew Reinhard}
\affiliation{Department of Physics, University of Florida,\\2001 Museum Road, Gainesville, FL 32611, USA}

\author{Elena Fernandez}
\affiliation{Instituto de Astrof\'{i}sica de Andaluc\'{i}a (CSIC),\\Glorieta de la Astronom\'{i}a, s/n, E-18008 Granada, Spain}

\author{Francisco Prada}
\affiliation{Instituto de Astrof\'{i}sica de Andaluc\'{i}a (CSIC),\\Glorieta de la Astronom\'{i}a, s/n, E-18008 Granada, Spain}

\author{Steven Ahlen}
\affiliation{Department of Physics, Boston University,\\590 Commonwealth Avenue, Boston, MA 02215, USA}

\author{Davide Bianchi}
\affiliation{Dipartimento di Fisica ``Aldo Pontremoli'', Universit\`a degli Studi di Milano,\\Via Celoria 16, I-20133 Milano, Italy}
\affiliation{INAF-Osservatorio Astronomico di Brera,\\Via Brera 28, 20122 Milano, Italy}

\author{David Brooks}
\affiliation{Department of Physics \& Astronomy, University College London,\\Gower Street, London, WC1E 6BT, UK}

\author{Todd Claybaugh}
\affiliation{Lawrence Berkeley National Laboratory,\\1~Cyclotron Road, Berkeley, CA 94720, USA}

\author{Axel de la Macorra}
\affiliation{Instituto de F\'{\i}sica, Universidad Nacional Aut\'{o}noma de M\'{e}xico,\\Circuito de la Investigaci\'{o}n Cient\'{\i}fica, Ciudad Universitaria, Cd.\ de M\'{e}xico  C.~P.~04510,  M\'{e}xico}

\author{Arnaud de Mattia}
\affiliation{IRFU, CEA, Universit\'{e} Paris-Saclay,\\F-91191 Gif-sur-Yvette, France}

\author{Biprateep Dey}
\affiliation{Department of Astronomy \& Astrophysics, University of Toronto,\\Toronto, ON M5S 3H4, Canada}
\affiliation{Department of Physics \& Astronomy, University of Pittsburgh,\\3941~O'Hara Street, Pittsburgh, PA 15260, USA}
\affiliation{Pittsburgh Particle Physics, Astrophysics, and Cosmology Center (PITT PACC), University of Pittsburgh,\\3941~O'Hara Street, Pittsburgh, PA 15260, USA}

\author{Peter Doel}
\affiliation{Department of Physics \& Astronomy, University College London,\\Gower Street, London, WC1E 6BT, UK}

\author{Enrique Gazta\~{n}aga}
\affiliation{Institut d'Estudis Espacials de Catalunya (IEEC),\\c/ Esteve Terradas 1, Edifici RDIT, Campus PMT-UPC, 08860 Castelldefels, Spain}
\affiliation{Institute of Cosmology and Gravitation, University of Portsmouth,\\Dennis Sciama Building, Portsmouth, PO1 3FX, UK}
\affiliation{Institute of Space Sciences, ICE-CSIC,\\Campus UAB, Carrer de Can Magrans s/n, 08913 Bellaterra, Barcelona, Spain}

\author{Gaston Gutierrez}
\affiliation{Fermi National Accelerator Laboratory,\\PO Box 500, Batavia, IL 60510, USA}

\author{Klaus Honscheid}
\affiliation{Center for Cosmology and AstroParticle Physics, The Ohio State University,\\191~West Woodruff Avenue, Columbus, OH 43210, USA}
\affiliation{Department of Physics, The Ohio State University,\\191~West Woodruff Avenue, Columbus, OH 43210, USA}

\author{Dragan Huterer}
\affiliation{Department of Physics, University of Michigan,\\450~Church Street, Ann Arbor, MI 48109, USA}

\author{Dick Joyce}
\affiliation{NSF NOIRLab, 950~N.\ Cherry Ave., Tucson, AZ 85719, USA}

\author{Robert Kehoe}
\affiliation{Department of Physics, Southern Methodist University,\\3215~Daniel Avenue, Dallas, TX 75275, USA}

\author{David Kirkby}
\affiliation{Department of Physics and Astronomy, University of California, Irvine, 92697, USA}

\author{Theodore Kisner}
\affiliation{Lawrence Berkeley National Laboratory,\\1~Cyclotron Road, Berkeley, CA 94720, USA}

\author{Martin Landriau}
\affiliation{Lawrence Berkeley National Laboratory,\\1~Cyclotron Road, Berkeley, CA 94720, USA}

\author{Laurent Le Guillou}
\affiliation{Sorbonne Universit\'{e}, CNRS/IN2P3, Laboratoire de Physique Nucl\'{e}aire et de Hautes Energies (LPNHE), FR-75005 Paris, France}

\author{Marc Manera}
\affiliation{Departament de F\'{i}sica, Serra H\'{u}nter, Universitat Aut\`{o}noma de Barcelona,\\08193 Bellaterra (Barcelona), Spain}
\affiliation{Institut de F\'{i}sica d’Altes Energies (IFAE), The Barcelona Institute of Science and Technology,\\Edifici Cn, Campus UAB, 08193, Bellaterra (Barcelona), Spain}

\author{Aaron Meisner}
\affiliation{NSF NOIRLab, 950~N.\ Cherry Ave., Tucson, AZ 85719, USA}

\author{Ramon Miquel}
\affiliation{Instituci\'{o} Catalana de Recerca i Estudis Avan\c{c}ats,\\Passeig de Llu\'{\i}s Companys, 23, 08010 Barcelona, Spain}
\affiliation{Institut de F\'{i}sica d’Altes Energies (IFAE), The Barcelona Institute of Science and Technology,\\Edifici Cn, Campus UAB, 08193, Bellaterra (Barcelona), Spain}

\author{Seshadri Nadathur}
\affiliation{Institute of Cosmology and Gravitation, University of Portsmouth,\\Dennis Sciama Building, Portsmouth, PO1 3FX, UK}

\author{Will Percival}
\affiliation{Department of Physics and Astronomy, University of Waterloo,\\200~University Ave.\ W, Waterloo, ON N2L 3G1, Canada}
\affiliation{Perimeter Institute for Theoretical Physics,\\31~Caroline St.\ North, Waterloo, ON N2L 2Y5, Canada}
\affiliation{Waterloo Centre for Astrophysics, University of Waterloo,\\200~University Ave.\ W, Waterloo, ON N2L 3G1, Canada}

\author{Ashley Ross}
\affiliation{Center for Cosmology and AstroParticle Physics, The Ohio State University,\\191~West Woodruff Avenue, Columbus, OH 43210, USA}
\affiliation{Department of Physics, The Ohio State University,\\191~West Woodruff Avenue, Columbus, OH 43210, USA}

\author{Eusebio Sanchez}
\affiliation{CIEMAT, Avenida Complutense 40, E-28040 Madrid, Spain}

\author{David Schlegel}
\affiliation{Lawrence Berkeley National Laboratory,\\1~Cyclotron Road, Berkeley, CA 94720, USA}

\author{Michael Schubnell}
\affiliation{Department of Physics, University of Michigan,\\450~Church Street, Ann Arbor, MI 48109, USA}

\author{Hee-Jong Seo}
\affiliation{Department of Physics \& Astronomy, Ohio University,\\139~University Terrace, Athens, OH 45701, USA}

\author{Joseph Silber}
\affiliation{Lawrence Berkeley National Laboratory,\\1~Cyclotron Road, Berkeley, CA 94720, USA}

\author{David Sprayberry}
\affiliation{NSF NOIRLab, 950~N.\ Cherry Ave., Tucson, AZ 85719, USA}

\author{Gregory Tarl\'e}
\affiliation{Department of Physics, University of Michigan,\\450~Church Street, Ann Arbor, MI 48109, USA}

\begin{abstract}
Here we report the first measurement of the parity-violating (PV) 4-Point Correlation Function (4PCF) of the Dark Energy Spectroscopic Instrument’s Year 1 Luminous Red Galaxy (DESI Y1 LRG) sample, motivated by the potential detection of the PV 4PCF in the Sloan Digital Sky Survey Baryon Oscillation Spectroscopic Survey (SDSS BOSS) galaxies. In our auto-correlation (``auto'') analysis, we find a statistically significant excess of the PV signal compared to mocks without any PV, at $\sim$4–10$\sigma$ depending on details of the analysis. This could arise either from genuine PV or from an underestimation of the variance in the mocks; it is unlikely to arise, at the signal level, from a systematic. We then cross-correlate (``cross'') the putative PV signal between different, independent patches of sky, and there find no detection of parity violation. The two measurements are in significant tension: while the cross has somewhat larger error bars than the auto, this is not sufficient to explain the discrepancy. We thus present the current work as an intriguing addition to the PV work on BOSS and as motivation for exploring further the relationship between the auto and cross PV 4PCF analyses.
\end{abstract}
						  
\maketitle

%%%%%%%%%%%%%%%%%%%%%%%%
\section{Introduction}
\label{sec:intro}

Searching for cosmic parity violation (PV) in the 4-Point Correlation Function (4PCF) of large-scale structure (LSS) was first proposed by \cite{cahn_short_pub} and then measured on SDSS BOSS data by the same group \cite{hou_parity_pub}, relying on techniques presented in \cite{iso_basis_pub}. \cite{hou_parity_pub} used the BOSS Constant stellar mass (CMASS) sample of Luminous Red Galaxies (LRGs; \num{777202} galaxies, $\bar{z} = 0.57$), finding evidence for PV at 4–7$\sigma$ depending on the analysis details. \cite{hou_parity_pub} also found some statistical evidence ($\sim$2–3$\sigma$) of PV in the smaller, lower-redshift LOWZ sample (\num{280067} galaxies, $\bar{z} = 0.32$). \cite{phil_parity} examined the same data set using similar methods and the covariance matrix taken from \cite{hou_parity_pub}. Its confirmation of \cite{hou_parity_pub} thus cannot really be deemed independent.\footnote{\cite{phil_parity} used a rank test, which can only describe the rarity of the data with respect to the (no-PV) \num{2000} BOSS mocks. Thus the maximum significance could only be one in \num{2000}, about $3\sigma$, which is indeed roughly what was found. In detail, the data had rank three of \num{2048} mocks, and so did not saturate this maximum, but there could always be undiagnosed issues with the two mocks that outrank the data, so this result is more robustly viewed as a lower bound. It also used a coarser binning (ten bins) than the main choice in \cite{hou_parity_pub}, and this leads to a lower-significance result for reasons discussed in \cite{hou_parity_pub} \S2.2 and \S5.1.4; \cite{hou_parity_pub} also tested ten bins and \cite{phil_parity}'s result is consistent with that.}
However, a later analysis explored whether these signals are the result of a mismatch in the \emph{parity-even} higher-order correlation functions between the used mocks compared to the data \cite{krol_parity}.

\textit{What is PV?}---A parity transformation ($\mathrm{P}$) is a spatial inversion, $(x,y,z) \to -(x,y,z)$; in 3D, it is equivalent to reflection through any 2D plane (``mirroring'') composed with a rotation. Since we take it that galaxy clustering is rotation-invariant (the cosmological assumption of isotropy), one can ignore the rotation and visualize parity transformation as simply \textit{mirroring}.\footnote{Wide-angle redshift-space distortions (RSD), which break the rotation invariance about each galaxy in the survey, and leave the only true rotation invariance as that about the observer, complicate this picture slightly. For instance, \cite{paul_prl, paul_longer} discuss this in the trispectrum (Fourier transform of the 4PCF). They do not find any parity-violation due to RSD after averaging over the line of sight.} $\mathrm{P}$'s effect on any 1D or 2D subspace can be reversed by a rotation in 3D. Hence, any statistic living in 1D or 2D that is averaged over rotations (such as the 2PCF or 3PCF) is \emph{not sensitive to parity}. 

%\textit{Understanding PV}---Now, PV would simply mean that there is a statistical imbalance between tetrahedra with a given set of six sides, and their mirror images, in a galaxy sample. Notably, specifying six side lengths of a tetrahedron is not enough---there are two ways to glue six sides into a tetrahedron, and the two resulting tetrahedra are mirror images of each other (Fig.~1 of \cite{hou_parity_pub}). To transform a tetrahedron into its mirror image, one can turn it inside out; in fact this is what spatial inversion, or P, means. Thus, if one considers the signed volume, mirroring flips its sign. Indeed, all of the P-odd basis functions \cite{iso_basis_pub} (which track these imbalances) have as their first factor the normalized, signed volume of the tetrahedron, given by the scalar triple product of three sides coming from a shared vertex \cite{hou_parity_pub} App. A, and \cite{iso_basis_pub} and \S5 of \cite{iso_gen_pub}\footnote{Tetrahedron volume:  \url{https://mathworld.wolfram.com/Tetrahedron.html}}.

\textit{Implications if Detected}---Now, if PV is of genuine cosmological origin, what would this mean? The forces of the Standard Model (SM) are all parity-conserving save for the weak force, which is too weak to impact galaxy clustering on cosmological scales \cite{cahn_short_pub}. Thus, if PV were genuine, it would have to indicate new physics during a regime when the SM is no longer sufficient---\textit{i.\,e.} in the high-energy, early-Universe epoch of inflation. Much work has explored this, considering PV from axions \cite{niu_axion, reinhard_axion, axion_4pt_cho}, inflation \cite{cabass_collider, cabass_no_go, thavanesan_no_go, lee_loops, stefanyszyn_factorization, jazayeri_nonlocality}, or lensing by chiral gravitational waves \cite{inomata_lensing}.
%Theoretical work has also explored 

%Theoretical work has also examined general criteria governing whether given inflationary scenarios and particle interactions can produce PV or not \cite{cabass_no_go, thavanesan_no_go}. \ak{suggest editing following sentence to: Parity violation could arise from loop corrections or non-locality in inflation.} Ideas such as leading loops, correlator factorization, and non-locality have also been done in the context of PV \cite{lee_loops, stefanyszyn_factorization, jazayeri_nonlocality}.

%Finally, an alternative proposal for explaining observed PV that also implies new physics, but at late times, is that there is no underlying PV in the galaxy distribution, but chiral gravitational waves (GWs) lens the LSS and thus our observations of them have PV \cite{inomata_lensing}. 

\textit{Importance of the 4PCF}---The lowest-order \emph{natively 3D} correlation function is the 4PCF, which probes the excess of tetrahedra over and above what a uniformly random spatial distribution of points would have. The 4PCF is thus the first choice to probe PV in 3D LSS \cite{cahn_short_pub}; \cite{shiraishi_16} noted that the harmonic-space trispectrum (projection of temperature anisotropies onto spherical harmonics whose arguments are the points' lines of sight to the observer) could be used in this way for the Cosmic Microwave Background (CMB).

Earlier work \cite{jeong_clus_fossils_pub} pointed out that pairs of ``local'' power spectra separated by some distance $\vec{x}$ could probe LSS PV---essentially, somewhat of a mixed-space (Fourier-position) 4-point function. \cite{jamieson_pop_pub} recently proposed ``Parity-Odd Power Spectra'' (POPS) to compress the odd sector and explore configurations for which the position-space 4PCF is less well-suited. POPS applies operators such as curl and gradient to the density field, then forms odd-parity combinations to integrate over some wave-vectors. POPS possess a more sparse and nearly diagonal covariance.

\textit{Numerous follow-up studies} have now been performed, including searches for evidence in the CMB \cite{phil_cmb}, in the Lyman-$\alpha$ forest \cite{adari_ly_alpha}, and in BOSS with a cross-correlation method designed to separate a genuine signal from mis-estimate of the covariance \cite{krol_parity};
all three returned null results.
%\ak{Do we want to say more about 24 (and more generally the fact that these follow-up searches did not detect PV)?}
%Zack---I want to criticize the Lyman alpha one a bit more because it wwas not good regarding covariance! I also want to comment that CMB primary is not good for this---we can cite Ale's and my trispectrum ppaper whcih explains that in detail.
\cite{phil_uchuu} explored using different mocks to assess the significance, but these mocks were made by replicating a smaller underlying box, and the impact of this on the covariance is not dealt with sufficiently to draw a clear conclusion.\footnote{Rescaling by $V_{\mathrm{mock}}/V_{\mathrm{underlying\;box}}$ should correct this effect, but relies on precise estimate of the mock volume. This was estimated using galaxies in \cite{ereza_uchuu}, but such an estimate is subject both to galaxy clustering (especially if the duplicated region is small) and shot noise. \cite{ereza_uchuu} App. B found this rescaling leaves a \SIrange{10}{15}{\percent} over-estimate of the covariance in the duplicated-box-mock unaccounted for in the 2PCF covariance. A rough scaling argument shows this would give a \SIrange{20}{30}{\percent} over-estimate in the 4PCF covariance, which explains the result of \cite{phil_uchuu}.} \cite{bao_odd} proposes searching for Baryon Acoustic Oscillations (BAO) in the odd 4PCF as a test of whether it is genuine. \cite{coulton_pv_sims} also simulated structure formation with PV initial conditions.

This work is structured as follows. We first outline our PV search \hyperref[sec:methods]{Methods}; then we describe our \hyperref[sec:data_cuts_mocks]{Data, Cuts, and Mocks}. We next discuss the \hyperref[sec:covariance]{Covariance}, then present \hyperref[sec:results]{Results}, and end with a \hyperref[sec:concs]{Concluding Discussion}. A reader eager to reach our key results quickly could focus on our \hyperref[sec:methods]{Methods}, \hyperref[sec:data_cuts_mocks]{Data, Cuts, and Mocks}, and \hyperref[sec:results]{Results}
sections upon initial reading and return to the other sections subsequently.

An \hyperref[sec:appendix]{Appendix} shows the footprint of the data and the sky cuts used for the cross analysis;
 the results from different analysis setups; examination of mock–data consistency; tests of the covariance; calibration of the covariance to different mocks; different ways of assessing detection significance; scaling comparisons for the variance of the statistics we use; and systematics exploration.

\section{Methods}
\label{sec:methods}

\textit{Auto, Compressed, and Cross}---We employ three PV search methods: the first is simply to measure the 4PCF parity-odd modes, and assess how significantly they deviate from zero (``auto''). 
In particular, we perform a template-free search---our ``model'' is that there is zero PV. We therefore compare the parity-odd four-point modes' summed squares, with the sum weighted by the inverse covariance (subscript ``ana'' for analytic), between data and mocks---this statistic is simply $\chi^2$ with the model set to zero, \textit{i.\,e.} 
\begin{align}
\label{eq:auto_def}
    \chi^2 \equiv \vec{\zeta}\, \mathbf{C}_{\mathrm{ana}}^{-1}\, \vec{\zeta}^{\,\mathrm{T}},
\end{align}
where $\vec{\zeta}$ is a vector of the 4PCF coefficients and $\mathrm{T}$ denotes transposition.

The second is to diagonalize the analytical covariance, select the highest precision (\textit{i.\,e.} smallest eigenvalue) several hundred eigenvectors, and then obtain the covariance of these linear combinations of our initial modes directly from mocks. We term this analysis ``compressed'' (\textit{e.\,g.} \S5.2.1 of \cite{hou_parity_pub}, and \cite{Scoccimarro_Data_Compression}). 
In this case the test statistic is formed in the same way as in the auto-analysis, but its values follow a $T^2$ distribution since both the covariance and data have noise, not just the data as in our auto method. We thus call the test statistic $T^2$ rather than $\chi^2$. 

The compressed method is conservative: it throws out information (corresponding to the eigenvectors that are not used), and so it will always give a lower bound on the true detection significance. This holds even if the theory covariance is not perfectly correct. In that case, one will have just not selected the true set of $N_{\mathrm{eig}}$ highest-precision eigenvectors, but rather some sub-optimal set.

Finally, we perform a ``cross'' analysis, by measuring all the odd modes on each spatially separated patch we have produced and then cross-correlating the results \cite{krol_parity}. We have
\begin{align}
\label{eq:cross_def}
    \chi^2_{\times} \equiv \frac{1}{N_{\mathrm{p}}(N_{\mathrm{p}} - 1)}\sum_{\mu \neq \nu} \vec{\zeta}^{(\nu)}\; \mathbf{C}^{-1}_{\mathrm{ana}}\; \vec{\zeta}^{(\mu) \, {\mathrm{T}}}, 
\end{align}
with $\vec{\zeta}^{(\mu)}$ a vector of the 4PCF coefficients on the $\mu^{\mathrm{th}}$ patch and $N_{\mathrm{p}}$ the number of patches; we highlight that this statistic cross-correlates only, so excludes the auto of any patch, where $\mu = \nu$; the pre-factor simply divides out the number of patch pairs contributing and so gives us an average.\footnote{If we take instead the sum over only $\mu = \nu$, and normalize by $1/N_{\mathrm{p}}$, we have $\chi^2$. Since both $\chi^2$ and $\chi^2_{\times}$ are thus averages over their estimates from the patches, the cross and auto are directly comparable; in particular the widths of their null distributions should be. This normalization equivalence is explained more extensively in \cite{krol_parity}}

Any genuine cosmological signal should (by homogeneity) be present in all patches and hence picked out by the cross. If the ``detection'' is actually due to under-estimation of errors, then that effect is random and will not appear in cross-corelation (since each region is statistically independent). 

The first method was proposed in \cite{cahn_short_pub} and first carried out in \cite{hou_parity_pub}. The idea of a cross-correlation of different regions was suggested and a cursory analysis performed in \cite{hou_parity_pub}. \cite{krol_parity} then showed formally that this approach could be used to separate a genuine signal from a false positive due to covariance mismatch between mocks and data. They carried out an extensive cross analysis, finding that the BOSS parity-odd detection was consistent with arising from covariance mismatch.

\textit{Maximum $\ell$ and Binning in Tetrahedron Side Length}---We measure the 4PCF projected onto all odd isotropic basis functions \cite{iso_basis_pub, iso_gen_pub} up to $\ell_{\mathrm{max}} = 5$.\footnote{The isotropic basis functions here expand the clustering of three galaxies (``secondaries'') around a fourth (``primary'') placed at the origin. The functions are products of one spherical harmonic for each of the secondaries, whose argument is the unit vector to that galaxy. The product of harmonics is weighted by a 3-$j$ symbol and summed over the orders $m_i$.} $\ell_{\mathrm{max}}$ controls the maximum degree of the spherical harmonics used to expand the angular dependence of each tetrahedron around a ``primary'' point, see \textit{e.\,g.} Fig.~1 of \cite{hou_parity_pub}. The basis functions that include one or more $\ell = 5$ are used solely for edge-correction \cite{se_3pt_alg, encore}, concisely reviewed in \S2 of \cite{hou_parity_pub}. In our analysis we then employ only those to $\ell_{\mathrm{max}} = 4$; there are 23 such odd functions, listed in \cite{hou_parity_pub} App.~A. We also measure the connected, even-parity 4PCF to the same $\ell_{\mathrm{max}}$, which involves another 42 basis functions. These results are discussed in the \hyperref[sec:appendix]{Appendix} and in \cite{hou_even_desi} (ten radial bins) and \cite{ortola_even_desi} (eighteen radial bins, matched to our setup in the present work). 

We bin each tetrahedron side extending from the primary vertex $(r_1, r_2, r_3)$ into 18 bins from \SIrange{20}{160}{\per\hHubble\Mpc}; this is very nearly the same choice as the finer binning scheme of \cite{hou_parity_pub}, with the reasons for a fine binning explained in \S2.2 there. We impose $r_1 < r_2 < r_3$ for reasons explained in \cite{iso_basis_pub} and we also enforce that $r_i - r_j \geq \SI{20}{\per\hHubble\Mpc}$ as this constraint ensures that all sides of all tetrahedra we explore exceed \SI{20}{\per\hHubble\Mpc}. Overall, we have \num{5060} degrees of freedom in our main analysis.

\textit{Algorithm}---The algorithm, developed based on the 3PCF works \cite{se_3pt_alg, se_3pcf_ft, garcia_aniso, se_3pcf_boss, se_3pcf_rv_data, se_rv_theory, se_3pcf_model}, exploits the analog of the spherical harmonic addition theorem but for the more-than-two-argument generalization of Legendre polynomials \cite{iso_basis_pub, iso_gen_pub}, to avoid ever looking at triples of galaxies around a given ``primary'' galaxy in the process of forming the 4PCF \cite{encore}. Rather, it forms outer products of sets of three harmonic coefficients $a_{\ell_i m_i}$ of the galaxy density field expanded on concentric spherical shells (radial bins). It then sums over the associated ``azimuthal quantum numbers'' ($m_i$) subject to a 3-$j$ symbol constraining the $\ell_i, m_i$ combinations, as isotropy and the Wigner–Eckart theorem dictate \cite{edmonds_57, luo_1994, spergel_goldberg_cmb, s_constrained_realization, zs_rvw}. 

\textit{In-practice scaling}---For the 4PCF, formation of this outer product and then summing over $m$ dominates the in-practice cost; since this must be done around each ``primary,'' the algorithm scales \textit{linearly} in the number of galaxies. We also measure the 2PCF and 3PCF, which in practice scale quadratically in number of galaxies, but the scalings' pre-factor is far less than the 4PCF's and so this latter part dominates the overall scaling.

\textit{Code \& Wall-Clock Computational Cost}---We use a proprietary GPU code, \textsc{cadenza}, to measure the 4PCF, based on the CPU code \textsc{encore} \cite{encore}, which in turn stems from the 3PCF code of \cite{se_3pt_alg}. For the 4PCF, \textsc{cadenza} is 140$\times$ faster than \textsc{encore} on the BOSS CMASS volume (comparable to DESI Year 1 (Y1) LRGs), and can obtain the 4PCF of data and 50$\times$ randoms in $\sim$45 minutes on one A100 GPU (the 2PCF takes of order 10 seconds, the 3PCF a few minutes). Thus the work reported here, on $\gtrsim$\num{1000} Y1-sized catalogs, could be done in a few thousand core-hours; with 100 GPUs on UF's HiPerGator system the entire pipeline runs start to finish in on the order of one day.

\section{Data, Cuts, and Mocks}
\label{sec:data_cuts_mocks}

\textit{Data}---The Dark Energy Spectroscopic Instrument (DESI) is currently conducting a five-year survey of \SI{14000}{\deg\squared} \cite{DESI_Survey_Operation_2023}, obtaining spectra of 40M galaxies and quasars to determine the nature of dark energy \cite{DESI2016a.Science, DESI2016b.Instr, DESI2022.KP1.Instr, DESI_spectroscopic_pipeline_2023, Corrector.Miller.2023, Poppett24, DESI_DR1_cosmology_2024, DESI_DR2_cosmology_2025, DESI_DR1_2025}. We use the DESI DR1 \cite{DESI_DR1_2025} LRG sample \cite{LRG.TS.Zhou.2023}, which extends from redshift \numrange{0.4}{1.1} with a nearly constant number density $n(z)$ of about \SI{5e-4}{\hHubble\cubed\per\Mpc\cubed} to $z=0.8$, and declining  thereafter \cite{KP3}. The DR1 LRG sample covers \SI{3650}{\deg\squared} in the North Galactic Cap (NGC) and \SI{2089}{\deg\squared} in the South Galactic Cap (SGC), and contains roughly 2.1M galaxies.

%ZS checking we have must-cite papers. August 5th 2025.
%We have all the Must-cite pappers listed here:
%https://desi.lbl.gov/trac/wiki/PubBoard/SupportRef
%There are 9 listed on the above and 9 listed below that have been checked.

%The DESI Experiment Part II: Instrument Design---yes
%Overview of the Instrumentation for the Dark Energy Spectroscopic Instrument---yes?
%Miller---yes
%Poppett---yes
%Guy---spec pipeline---yes
%Schlafly---survey ops---yes
%KP2---DESI DR1 2025---yes
%KP7---DESI DR1 Cosmology 2024---yes
%DESI DR2 Cosmology---yes.

\textit{Cuts: Full Sample, Regions, and Patches}---We look at the \textit{full} Y1 LRG data-set as above. We also define two (spatially separated) \textit{regions} in the North and one in the South based on high completeness; we restrict to $0.4 < z < 0.8$ to remain in the flat part of $n(z)$. We always refer to these as \textit{regions}. We then also cut our full data into four \textit{patches} in the NGC and three in the SGC (\cref{fig:footprint}). We do so to produce spatially well-separated patches that can be considered statistically independent and used for our cross analysis \cite{krol_parity}. Their footprints are set based on desiring roughly equal area and high completeness, and here we use the full redshift range, $0.4 < z < 1.1$. In the main text, we present the auto analysis on the full sample and the cross analysis on the patches; the patch and region auto analyses are in the \hyperref[sec:appendix]{Appendix}.
%ZS above is pretty good after editing, July 27 2025. Could have one more polish round.

\textit{Mocks}---Following the DESI DR1 analyses, we use two sets of mocks, \num{1000} \textsc{EZMock}s \cite{Chuang15} and 25 \textsc{Abacus} mocks \cite{Maksimova21, Garrison21}. The former are produced using the Zeldovich approximation plus simple prescriptions for galaxy biasing, while the latter are $N$-body simulations. Both are tuned to match the 2-point clustering of the data, with galaxies populated using the \textsc{AbacusHOD} code \cite{Yuan22} for the \textsc{SecondGen} mocks \cite{KP3}. 

The \textsc{Abacus} mocks take a box smaller than the DESI volume (\SI{2}{\per\hHubble\cubed\Gpc\cubed}) and replicate and cut it as appropriate to match the survey geometry. This replication means the number of independent Fourier modes in the \textsc{Abacus} mocks does not match that of the data. The variance of the measured \textsc{Abacus} 4PCF, when assessed against the covariance calibrated to \textsc{EZMock}s, will also be higher for this reason. 

To correct for the \textsc{Abacus} volume replication, we apply the volume scaling of the covariance matrix, multiplying by the ratio of the full, replicated volume of the mock, $V_{\mathrm{mock}}$, to the unique volume, $V_{\mathrm{unique}}$.\footnote{We also apply a small additional scaling to account for the non-Gaussian $V^{1.5}$ scaling of the error on $\chi^2$, following App.~B.1 of \cite{krol_parity}.} This factor, $V_{\mathrm{mock}}/V_{\mathrm{unique}}$, is 1.25 (1.15) for NGC (SGC).

Two methods of fiber assignment are applied to the \textsc{Abacus} mocks: Alternative Merged Target Ledger (AltMTL) \cite{Lasker25}, which uses nearly the same fiber-assignment algorithm as the data, and Fast Fiber Assign (FFA) \cite{KP3s11-Sikandar}, which uses a much cheaper approximate procedure. Only FFA was run for \textsc{EZMock} as AltMTL was too expensive.

We use \textsc{Abacus} AltMTL as our baseline mocks, as they have the most realistic dynamics and fiber-assignment method. In particular, it was found in the DR1 full-shape analysis that the \textsc{EZMock} 2PCF covariances require an empirical scaling to match the more accurate \textsc{RascalC} covariance matrices \cite{Rashkovetskyi24} (as described in \S 5.7 and Table~3 of \cite{DESIFS}). This
suggests the possibility of a mismatch between the higher-point functions of \textsc{EZMock} FFA and the data. As a result, we disfavor comparing our results to \textsc{EZMock} FFA, even though these larger mocks do not require a correction for replicated volume.

%We correct for this volume mismatch when using \textsc{Abacus}
%We use a simple prescription to correct for these missing modes

%Hence the \textsc{Abacus} mocks must have a replication correction before being used to either to obtain the width of a null no-PV distribution, or to calibrate a covariance. We determine how to perform such a correction and do so, as we will discuss later. We also create mock sub-samples with cuts matching all those imposed on the data.

\begin{figure*}
    \centering
    \includegraphics[width=0.48\linewidth]{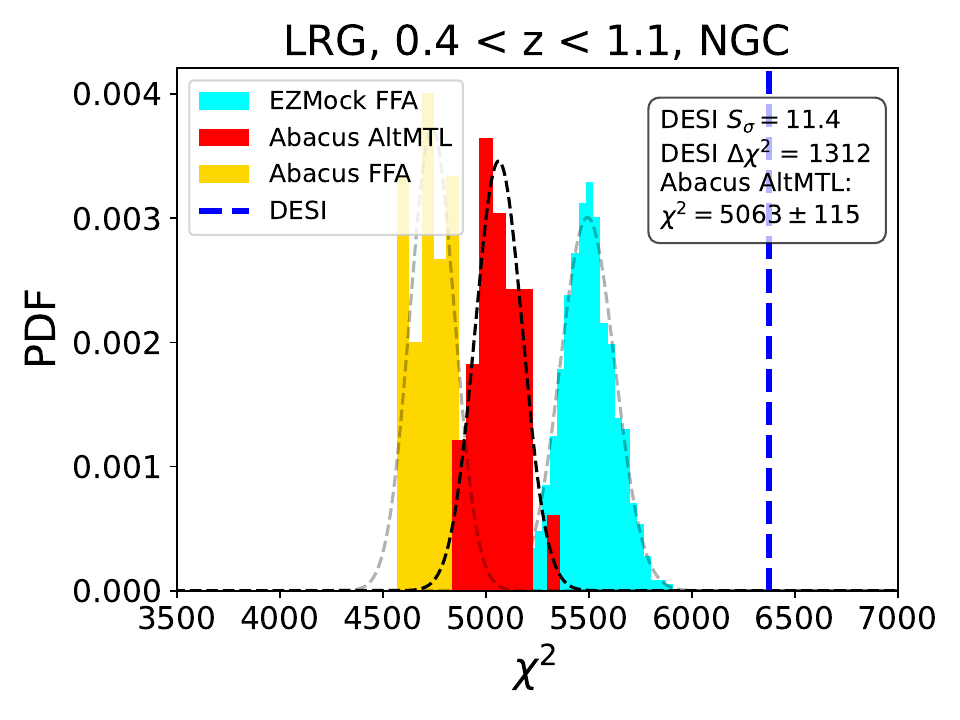}%
    \hfill%
    \includegraphics[width=0.48\linewidth]{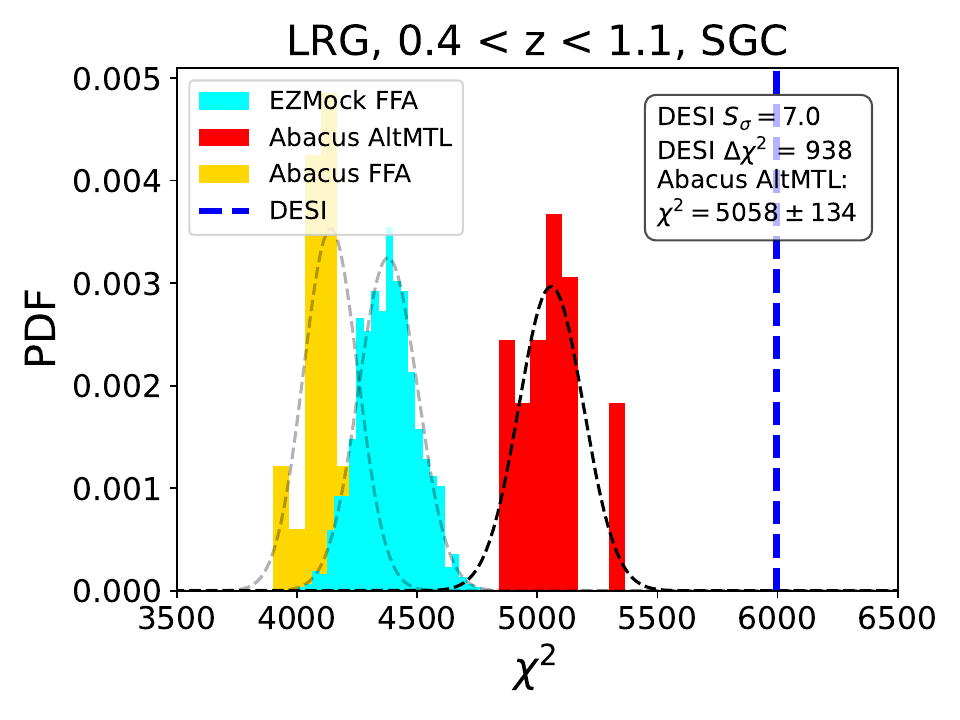}

    \caption{%
    Auto analysis for PV using $\chi^2$ (\pcref{eq:auto_def}) for NGC (left) and SGC (right); there are \num{5060} d.o.f. 
    Yellow is 25 \textsc{Abacus} mocks with Fast Fiber Assign (an approximate scheme; FFA), cyan is \num{1000} \textsc{EZMocks} with FFA, and red is \textsc{Abacus} with the more precise Alternative Merged Target Ledger (AltMTL) fiber assignment method. Here we have used the analytic covariance calibrated to \textsc{Abacus} AltMTL as described in the main text and summarized in Table~1, and both \textsc{Abacus} results have had replication correction applied also as described there. We fit a Gaussian to each distribution, 
    define $\Delta \chi^2$ as the DESI (blue dashed line) data's distance from the mean of \textsc{Abacus} AltMTL, and define the significance $S_\sigma$ by dividing $\Delta \chi^2$ by the Gaussian's width.
    We compare to \textsc{Abacus} AltMTL as our baseline mocks because they come from the most realistic simulation and fiber-assignment algorithm. We show the other mocks for reference; since the covariance was not calibrated to them, their $\chi^2$ distributions need not center at the number of d.o.f.}
    \label{fig:auto-full-ngc-sgc}
    %method-type-n or s.
\end{figure*}

\begin{figure*}
    \centering
    \includegraphics[width=0.48\linewidth]{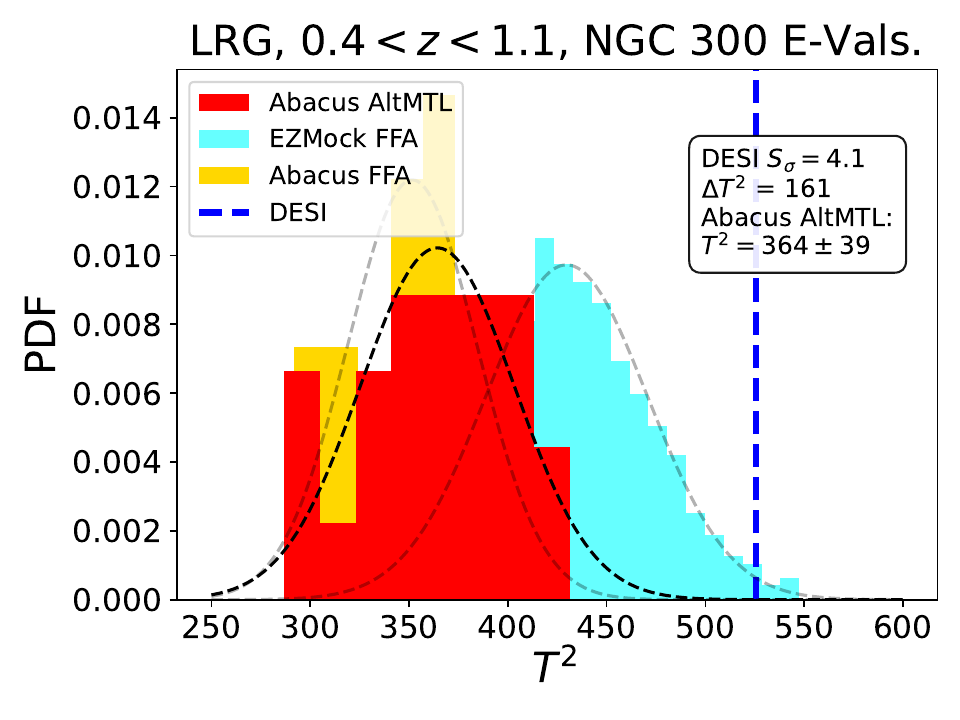}%
    \hfill%
    \includegraphics[width=0.48\linewidth]{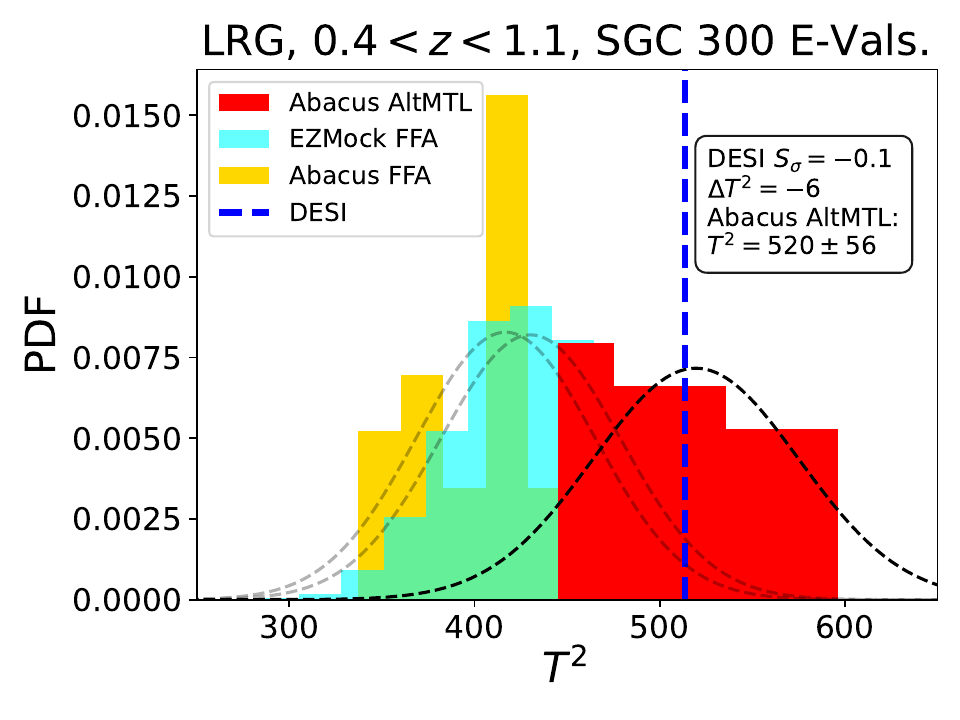}

    \caption{Here we show the compressed analysis 
    using a covariance matrix computed from the 300 lowest-noise (as dictated by the analytic covariance, calibrated to \textsc{Abacus} altMTL) eigenvectors; these eigenvectors then have their covariance measured from the \textsc{EZMocks}. We see 4.1$\sigma$ (NGC) and $-0.1\sigma$ (SGC) deviation from mocks when assessing distance from the center of the \textsc{Abacus} AltMTL distribution in units of its width, obtained by fitting a Gaussian (black dashed curve) to this distribution. Here our test statistic is $T^2$ because it is computed in the same way as $\chi^2$ but with an empirical covariance matrix, as further discussed in the main text. %The $T^2$ distribution with 300 d.o.f. and a covariance computed from \num{1000} mocks peaks at 429, and its standard deviation is 42. We ascribe the higher $T^2$ values of the mocks in SGC to its lower completeness than NGC, which the mocks reflect as they have had fiber-assign applied.
    %\ak{I don't agree with this sentence. The $T^2$ distributions of both NGC and SGC mocks are reasonably consistent with the theoretical distribution for 1000 mocks and 300 eigenvalues. The SGC mock distribution is 1.6$\sigma$ high vs.\ theory, and the NGC mock distribution is 1.7$\sigma$ low. The errors are similar (the theory error is 41). This does mean that if I used the theory distributions instead of the mocks, I would get 2.3$\sigma$ in NGC and 2.0$\sigma$ in SGC. I would interpret this as no detection in either case (I don't believe in $<3\sigma$!) and consistency between the two hemispheres. The other eigenvalues are similar? check!}
    The significance in Gaussian $sigma$, $S_{\sigma}$, is computed as $\Delta T^2/$[error bar (width)] of the \textsc{abacus} AltMTL distribution, \textit{i.\,e.} $161/39\sigma = 4.1\sigma$ for NGC and $-6/56\sigma = -0.1\sigma$ for SGC.}
    \label{fig:compresse-ngc-300}
    %method-type-n or s.
\end{figure*}

\begin{figure*}
    \centering
    \includegraphics[width=0.48\linewidth]{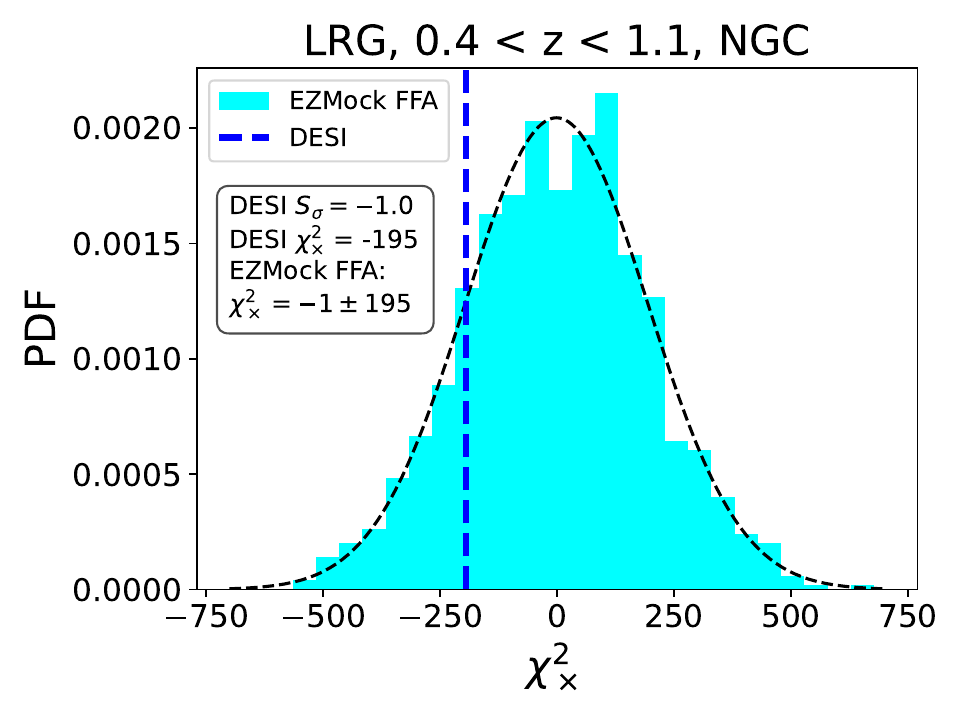}%
    \hfill%
    \includegraphics[width=0.48\linewidth]{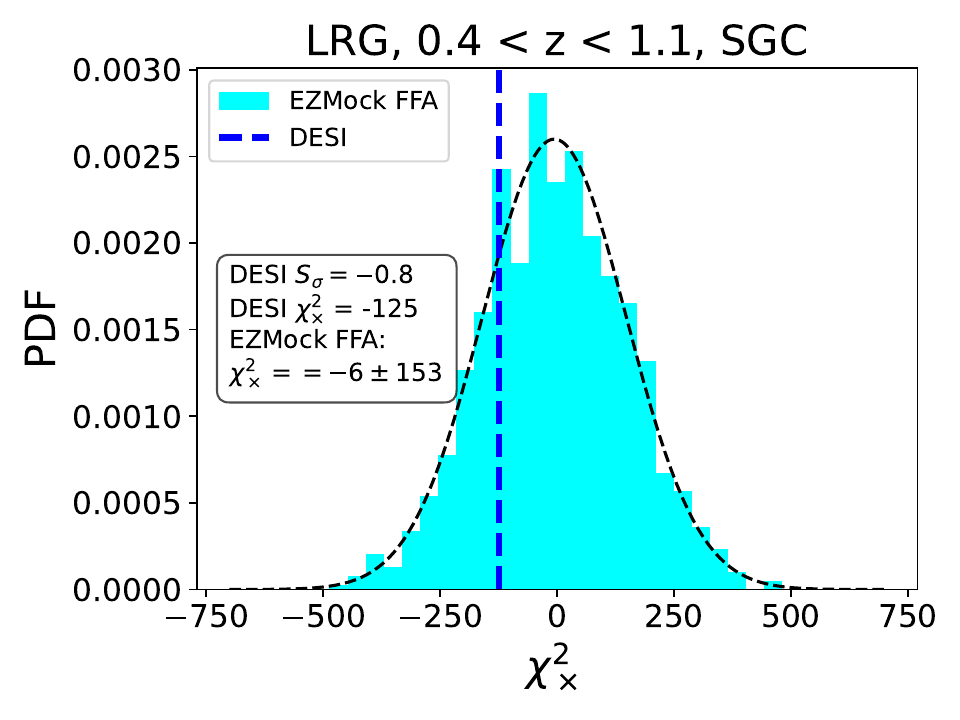}

    \caption{Cross-correlation (``cross'') results in  NGC (left) and SGC (right). We cross-correlate
    the PV data vector in four roughly equal patches in NGC and three roughly equal patches in SGC (shown in \cref{fig:footprint}). We compare to the cross in \textsc{EZMocks} (cyan), which have no PV. The cross is performed in the eigenspace of the analytic covariance, as further detailed in \cite{krol_parity} and \cref{eq:cross_def}. The dotted curves are Gaussian fits to the results. The cross does not give evidence for PV in DESI, and it is normalized so that the size of the error bars (standard deviation of the fitted Gaussian) is directly comparable to that of the auto (see \pcref{eq:cross_def} and the footnote below it). The error bars of $\chi^2_\times$ are 2.1$\times$ larger in NGC and 1.3$\times$ larger in SGC than the error bars of the auto $\chi^2$. For a detection of PV consistent with the auto and assuming statistical isotropy, the blue line would need to deviate $\sim$6-7$\sigma$ from the mocks' distribution and in the positive direction.} 
    \label{fig:combined_cross}
\end{figure*}

\section{Covariance Matrix}
\label{sec:covariance}

The covariance matrix measures the independence of each 4PCF mode, and radial triple-bin within that mode (\textit{e.\,g.} \cite{cahn_short_pub, encore}), from the others. Our ``auto'', ``compressed'', and ``cross'' approaches all share at some level the use of an analytic  covariance $\mathbf{C}_{\mathrm{ana}}$ computed by assuming a Gaussian Random Field (GRF) density, \textit{i.\,e.} to construct a detection significance $\chi^2$ as in \cref{eq:auto_def}.
%of the form $\vec{\zeta} \mathbf{C}_{\mathrm{ana}}^{-1} \vec{\zeta}^{\mathrm{T}}$, with $\vec{\zeta}$ a vector of the odd 4PCF coefficients in the isotropic basis and subscript ana denoting analytic covariance. 
The GRF covariance neglects higher-order correlations in the density field, which are of course present, but it does give the leading terms. Details are given in \cite{hou2021analytic}, extending the method of \cite{se_3pt_alg, xu_2012} for the 3PCF and 2PCF covariances, and with discussion of higher-order corrections in \cite{ortola_cov_I, ortola_cov_II}. Fuller discussion of the analytic covariance as applied to PV with the 4PCF is in \S4.2 of \cite{hou_parity_pub} and App.~A of \cite{krol_parity}.\footnote{We use the public \textsc{Julia} code \textsc{Analytic4PC} from \cite{krol_parity}: \url{https://gitlab.com/Socob/Analytic4PC}, which was tested against (and in agreement with) the \textsc{GlenCadaM} code of \cite{hou2021analytic}, \url{https://github.com/Moctobers/npcf_cov}.}

\textit{Analytic Covariance Inputs}---Briefly, the analytic covariance takes as an input the galaxy power spectrum, $P$, as well as a number density $\bar{n}$, combined as $P + 1/\bar{n}$. The second term represents the Poisson noise (shot noise) due to the discreteness of the galaxies.\footnote{Since $\bar{n}$ is a free parameter for which we end up fitting, it could in principle capture sub-Poissonian shot noise due to the halo exclusion effect discussed in \cite{ginzburg_exclusion, baldauf_exclusion, paech_exclusion}.}

\textit{Power Spectrum}---For $P(k)$ (shown in \cref{fig:template}), we use a \textsc{velocileptors} \cite{Chen20, Chen21} Effective Field Theory (EFT) of LSS model fit to the galaxy power spectrum of our sample. It differs from the official DESI full-shape analysis \cite{DESIFS} in that we do not apply the $\theta$-cut correction for fiber-assignment used there \cite{Pinon25}. We omit the $\theta$-cut since we cannot do it for the 4PCF, and we want the power spectrum used in the covariance to be on the same data and weights as those on which our 4PCF is measured.

\textit{Number Density}---$\bar{n}$ is an effective number density that is optimized to make the inverse of the analytic covariance come as close as possible to yielding the identity matrix when multiplied by the (noisy, and hence non-invertible) empirical covariance matrix. This latter is estimated from measuring the 4PCF on the \textsc{EZMocks}. In practice, we maximize the likelihood $\mathcal{L}$ given by Eq.~(55) of \cite{hou2021analytic} with respect to $\bar{n}$ and volume $V_{\mathrm{eff}}$ (see below):
\begin{align}
    \label{eq:covar_likelihood}
    - \log\mleft(\mathcal{L}(\bar{n}, V_{\mathrm{eff}})\mright) &\propto \mathrm{Tr} (\mathbf{C}^{-1}_{\mathrm{model}}(\bar{n}, V_{\mathrm{eff}})\mathbf{C}_{\mathrm{mocks}}) \,
    \\
    &\qquad -\log(\det) \, \mathbf{C}^{-1}_{\mathrm{model}}(\bar{n}, V_{\mathrm{eff}}).
    \nonumber
\end{align} 
We explored $\bar{n}$ from \SIrange{0.5e-4}{3e-4}{\hHubble\cubed\per\Mpc\cubed} in steps of \SI{0.1e-4}{\hHubble\cubed\per\Mpc\cubed}, using at each $\bar{n}$ the optimal volume $V_{\mathrm{eff}}$ for that $\bar{n}$, which can be determined analytically. We do the optimization separately for NGC and SGC, since they have quite different levels of completeness (see Fig.~2 in \cite{KP3}) and therefore likely different $N$-point correlation function (NPCF) variance. However, when we consider the different patches or regions within NGC and SGC, we do not re-optimize the covariance for each patch or region, but instead use the same NGC or SGC covariance everywhere.\footnote{We tested re-optimizing the covariance for each patch separately, and found this made almost no difference (and the per-patch optimized number densities were very similar to the overall optimal number density).}

\textit{Volume}---The analytic covariance also requires an effective volume, $V_{\mathrm{eff}}$. In principle, it is not exactly the same as the effective volume computed as a weighted integral of powers of $\bar{n}(z)$ over the survey (\textit{e.\,g.} \cite{de_Putter_2012, wadekar_2020}, \cite{hou2021analytic} Eq.~53), but in practice it is often numerically similar. We show the optimized $\bar{n}$ and $V_{\textrm{eff}}$ in \cref{tab:optimal_nbar_volume} in the \hyperref[sec:appendix]{Appendix}.

In our baseline results, we optimize the parameters of the analytic covariance to match the distribution from the \textsc{Abacus} AltMTL mocks, which have the most realistic dynamics and fiber-assignment implementation. The reduced volume of the \textsc{Abacus} mocks due to box replication means that the $V_{\mathrm{eff}}$ derived from them is smaller than the true volume of the data by $V_{\mathrm{cutsky}}/V_{\mathrm{unique}}$. Consequently, we scale the optimal $V_{\mathrm{eff}}$ by this replication factor. 

After doing this, the \textsc{Abacus} mocks themselves will then have larger fluctuations than would be expected for this scaled volume, since they are of a smaller unique volume. Thus to compare fairly the \textsc{Abacus} AltMTL and FFA mocks to data, we divide their $\chi^2$ by the replication factor. No such correction is needed for the data or \textsc{EZMocks}, since we are using the correct $V_{\mathrm{eff}}$ for an unreplicated volume in the covariance. For the compressed analysis, no correction factor is needed for the covariance, as it is determined from the un-replicated \textsc{EZMocks}, but we do divide $T^2$ from \textsc{Abacus}  by the replication factor to correct for the increased variance from the smaller effective volume.

\textit{Covariance Tests}---We test our analytic covariance using the ``half-inverse'' test, where we compute $\mathbf{C}_{\mathrm{ana}}^{-1/2} \mathbf{C}_{\mathrm{mocks}} \mathbf{C}_{\mathrm{ana}}^{-1/2} - \mathbf{I}$; if the analytic covariance were perfect, this would be a matrix of Gaussian noise with standard deviation $1/\sqrt{N_{\mathrm{mocks}}}$. This test is shown, and further discussed, in the \hyperref[sec:appendix]{Appendix} text and \cref{fig:covariance_test} there.

\textit{Covariance for Compressed Analysis}---The ``compressed'' analysis uses the analytic covariance as a starting point: we diagonalize it, and then choose the $N_{\mathrm{eig}}$ eigenvectors  with the smallest eigenvalues (highest precision) \cite{Scoccimarro_Data_Compression}. We then obtain an 
\textit{empirical} covariance matrix measured from the \num{1000} \textsc{EZMocks}. This is possible as long as $N_{\mathrm{eig}} \ll N_{\mathrm{mocks}}$. This approach was also used in the even-parity 4PCF BOSS analysis of \cite{even-4pcf}.

\section{Results}
\label{sec:results}

As outlined above, we explore three ways of slicing the data: the full Y1 sample; the sample cut into seven equal-area, high-completeness, spatially-separated \textit{patches} on the full $z$ range; and finally three \textit{regions} (two NGC, one SGC) cut to $z < 0.8$ and with footprint set by demanding high completeness.

We first discuss mock–data consistency, next our auto $(\chi^2)$ results, then our compressed $(T^2)$ analysis, and finally our cross $(\chi_{\times}^2)$ analysis.

\subsection{Mock–data Consistency}

As a check of the agreement between data and mocks in the even-parity sector, we compute the 2PCF, 3PCF, and even, connected 4PCF of both and compare. This matters because we use our mocks to calibrate the covariance (and then, in the compressed analysis, to form it). Hence, we want the mocks to mirror the data as well as possible in the even sector, since the odd 4PCF covariance will include contributions from the 2PCF, 3PCF, and even 4PCF in principle (though in practice this all must be absorbed by the $1/\bar{n}$ piece of our GRF analytic covariance). 

However, because NPCFs with $N>4$ also enter the covariance, the good agreement between the data and mock 2PCF, 3PCF, and even-parity 4PCFs does not foreclose the possibility of a mock–data mismatch in $\chi^2$. We show representative plots of our 2PCF, 3PCF, and even-parity 4PCF consistency tests in the \hyperref[sec:appendix]{Appendix}, \cref{fig:2_3_T2,fig:even_4_T2}. Overall, we find good consistency. We now move to our main results.

\subsection{\boldmath Auto Analysis: \texorpdfstring{$\chi^2$}{χ²}}

Comparing to the most faithful mock suite---the 25 \textsc{Abacus} AltMTL mocks---the auto analysis gives a 11.4$\sigma$ signal in NGC and 7.0$\sigma$ in SGC (\cref{fig:auto-full-ngc-sgc}).
The three mock distributions (\textsc{Abacus} FFA and AltMTL, \textsc{EZMock} FFA) have different centers and slightly different widths because of the different methods of fiber-assignment (FFA vs.\ AltMTL) on the same mock, and also differences in the small-scale behavior of the mock, including $N$-body vs. approximate dynamics and the galaxy–halo connection (\textsc{Abacus} vs.\ \textsc{EZMock}).

The detection significance is sensitive to the correction we applied to deal with the replications in \textsc{Abacus}. Here we use the simple scaling of $V_\mathrm{cutsky}/V_{\mathrm{unique}}$; deviations from it would change our significance when assessed with respect to \textsc{Abacus}.

On the other hand, comparing to \textsc{EZMock} Fast Fiber Assign (FFA; an approximate fiber assign method), which was produced without replication, gives statistical evidence of 6.6$\sigma$ in NGC and 5.7$\sigma$ in SGC. In the \hyperref[sec:appendix]{Appendix}, we show the analogous results for the \textit{regions} (\cref{fig:auto_regions_all_three}) and \textit{patches} (\cref{fig:auto_patches_ngc_sgc}).
We also show that the results are robust to removing imaging systematic and redshift failure weights (Table~\ref{tab:widths_zfail_imsys}), and explore the impact of fiber assignment on the auto (\cref{fig:complete_vs_altmlt_vs_FFA}).

%We show the results of combining the NGC and SGC patches also used for the cross analysis. We see that the \textsc{EZMock} and \textsc{Abacus} distributions have similar centers and similar widths; the data $\chi^2$ is very well sepparated from both. Since the \textsc{Abacus} have more faithful fiber assign and also likely more faithful dynamics, we use their distribution's width to assess significance. We find $12.5\sigma$.

%We show these plots for the NGC and SGC separately in the \hyperref[sec:appendix]{Appendix}. 

\subsection{Compressed Analysis}

We find lower statistical significance with the compressed analysis, of $\sim 4\sigma$ when adding NGC and SGC in quadrature (\cref{fig:compresse-ngc-300}), and when restricting to the 300 lowest-noise eigenvalues. This implies that the signal is spread out among many eigenvalues, which will be inaccessible to our compressed analysis because the total number of degrees of freedom (\num{5060}) exceeds the available number of \textsc{EZMock}s (\num{1000}). 
We use \textsc{EZMock} FFA to construct the covariance, as the 25 \textsc{Abacus} mocks available do not permit us to use enough eigenvalues. We then assess significance by comparing the data $T^2$ to the mean and width of the \textsc{Abacus} AltMTL. While the mock distribution is broadly consistent with a $T^2$ distribution (which would have mean 430 and standard deviation 42 with 300 eigenvalues), there are small but significant deviations in the mean. We view the comparison to the mock distribution as the more conservative choice, since the assumptions underlying the validity of the $T^2$ distribution may be violated. If we instead compared to the analytic $T^2$ distribution and converted the probability-to-exceed (PTE) to an equivalent number of Gaussian $\sigma$, we would obtain evidence of 2.1$\sigma$ (1.9$\sigma$) for NGC (SGC), or 2.8$\sigma$ combined.

We also explored other numbers of eigenvalues (50, 100, 150, 500) and found that the results are broadly consistent, though with a slightly smaller significance when using $<300$ eigenvalues and a slightly larger significance when using more, as generically expected if the signal is distributed among the eigenvalues. We also tested our compressed analysis by using the analytic covariance, checking that as we went to the full number of eigenvalues we recovered the auto analysis significance.

\subsection{\boldmath Cross Analysis: \texorpdfstring{$\chi^2_{\times}$}{χ²-×}}

In contrast to the auto, the cross analysis does not find any statistical evidence for PV ($-1.0\sigma$ and $-0.8\sigma$ detection in NGC and SGC). The \textsc{Abacus} boxes are unsuitable for the cross statistic---the PV 4PCF measurements (which are just noise, as \textsc{Abacus} has no real PV) in the different patches are correlated by the replication and therefore PV would be spuriously ``detected'' in them. Consequently we use the \textsc{EZMocks} instead.\footnote{If we remove \textsc{Abacus} patch pairs with a significant correlation, using four instead of six patch pairs in NGC, we also find no detection of PV but with a \SI{30}{\percent} larger error bar.}

The statistics are normalized so that $\chi^2_\times$ and $\chi^2$ are directly comparable \cite{krol_parity}. Comparing the widths of the mocks' distributions in \cref{fig:auto-full-ngc-sgc,fig:combined_cross}, we see that the cross statistic is 2.1$\times$ less constraining than the auto in NGC and \SI{30}{\percent} less constraining in the SGC. If the cross and the auto were perfectly consistent, we would expect a 6.7$\sigma$ (6.1$\sigma$) detection in the cross in NGC (SGC), derived by dividing $\Delta \chi^2$ from \cref{fig:auto-full-ngc-sgc} by the width of $\chi^2_\times$ in \cref{fig:cross_combined}. Thus, the lack of detection in the cross implies a significant tension between the auto and cross.

\section{Concluding Discussion}
\label{sec:concs}

Using DESI Y1 LRGs, we have a signal in the auto analysis, at 4–10$\sigma$, when comparing data to no-PV mocks. However, the non-detection of the signal in the cross analysis raises the possibility that the signal comes from the mocks underestimating the true variance of the data.
%the interpretation of this signal is not yet clear; in this template-free detection method, it could come from genuine PV or from underestimation by the mocks of the true variance of the data. 
As our \hyperref[sec:appendix]{Appendix} outlines, it is not likely to come from a systematic that produces a true PV signal in the infinite-volume-average limit, but there are systematics that could alter the variance, which we have carefully explored.

We do not detect PV when cross-correlating the signal in different patches on the sky. A detection in the cross analysis would be expected if the auto signal were coming from PV rather than mocks' underestimating the data's variance.

The situation is thus similar to what was found in BOSS, though with a different dataset and using higher-fidelity $N$-body mocks as our null case, which are state of the art in terms of estimating the variance. Future work should thus explore the origins of this discrepancy, \textit{e.\,g.} by measuring the sensitivity of the parity-odd evidence to the details of the galaxy–halo connection.

% Comparing the auto and cross, we see a difference---we have strong statistical evidence in the auto, but not in the cross. Up to the coverage fraction $f_{\mathrm{cov}}$ (the cross loses some area relative to the auto, in order to keep the patches separate), the signal is the same in each method. However, the variance of each statistic is different; in our case, the cross has slightly higher variance. Using these considerations, one can show that the detection significance ratio should be \textbf{ZS: is below the right normalization convention regarding Np? wonder if that factor should go away.}
% \begin{align}
%     S_{\times} / S_{\mathrm{auto}} =  f \sqrt{ (N_{\mathrm{p}}/(N_{\mathrm{p}} - 1) \mathrm{Var}(\chi^2) / \mathrm{Var}\chi^2_{\times})}
% \end{align}
% Using the definition in the \hyperref[sec:appendix]{Appendix} to compute $f_{\mathrm{cov}}$ and reading off variances from the relevant Figures, we see that the cross should find about \SI{80}{\percent} the detection of the auto for our sample. \textbf{ZS: this statement depends on which sample we compare---for instance, full? and it would be affected by the replication correction we need to do, which probably leads to underestimating the variance of auto and thus would lead to under-estimating this fraction.}

%\textbf{ZS: Alex I think DESI has some required acks. I can add them.}

\appendix

\section*{Appendix}
\label{sec:appendix}
%checklist as of July 20th 2025, atlanta. 
%patch and region footprints---done
%auto-combined sgc patches---done
%auto-full-ngc---done
%auto-full-sgc---done
%auto-combined-regions---done but needs retitling to ouse R1, R2, etc. not CT1, etc.
%auto-region-1---done but needs retitling
%auto-region-2---done but needs retitling
%auto-region-3---done but needs retitling

%cross
%cross-ngc---done
%cross-sgc---done

%right now, cross combined is in main text. this is not consistent with the auto choice  where we showed s and n separately, but that choice is dictated by Abacus. 

%covariance calibration
%ngc with 3---done but needs darker colors on cyan and yellow
%sgc with 3---done but same as above.

%missing from main text
%compressed one 

%missing from appendix
%covariance half-inverse test plot
%variance scaling plot with error bars--2 options it seems
%Pk we used for covariance matrix 

%missig from supplemental not reviewed material or appendix
%3PCF, even 4PCF

\begin{figure*}
    \centering
    \includegraphics[width=0.48\linewidth]{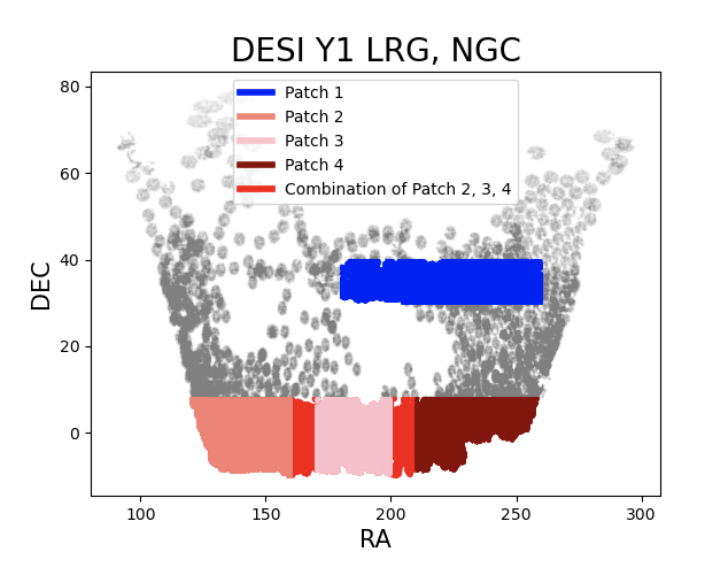}%
    \hfill%
    \includegraphics[width=0.48\linewidth]{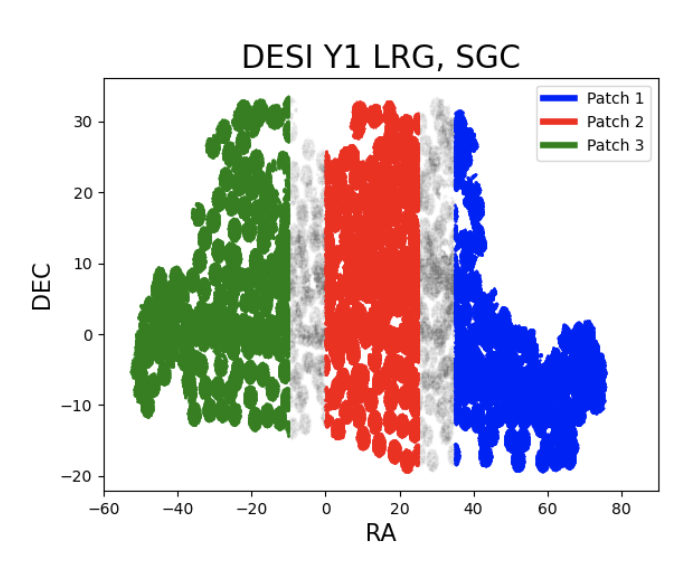}

    \caption{DESI Y1 footprints in the NGC (left) and SGC (right) showing the full sample (gray), regions, and patches. Patches 2, 3, and 4 in NGC are separated by the two vertical scarlet stripes, so that we have 4 patches in the NGC and 3 in the SGC. We define two regions in the NGC (upper, R1, in blue and lower, R2, being Patch 2, 3, 4 plus the scarlet stripes) and one in the SGC (its entirety). All \textit{regions} are cut to $0.4 < z < 0.8$, as motivated by $n(z)$'s flatness to that $z$ (\cref{fig:template}), while the \textit{patches} contain the entire redshift range, $0.4 < z < 1.1$.}
    \label{fig:footprint}
\end{figure*}

\textit{Footprints}---We first show the footprints of the full sample, our patches, and our regions (\cref{fig:footprint}). 

\begin{figure*}
    \centering
    \includegraphics[width=0.48\linewidth]{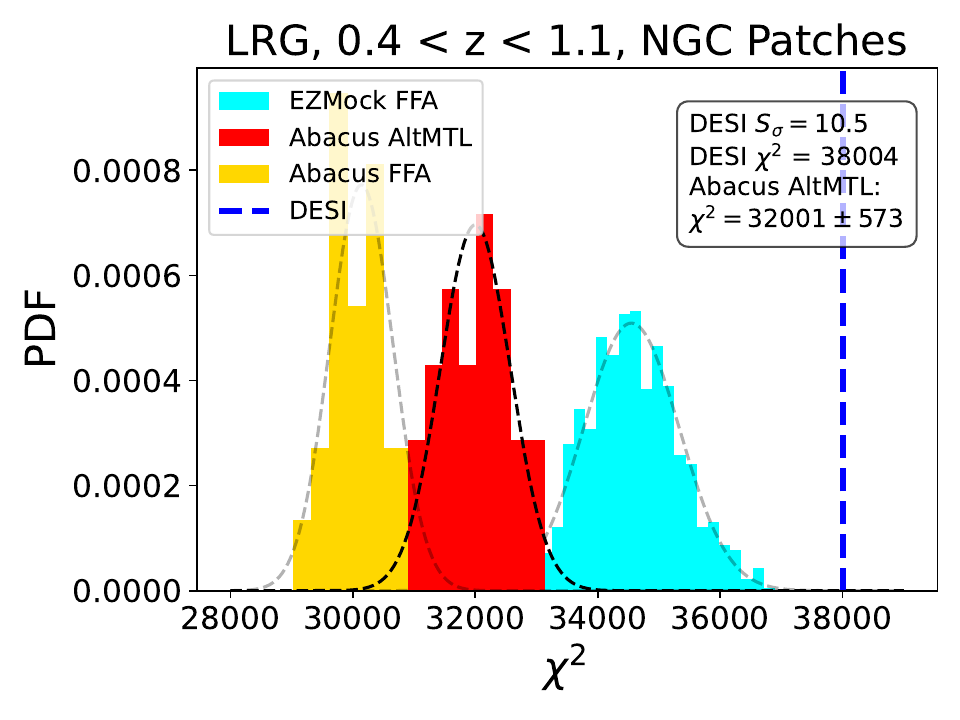}%
    \hfill%
    \includegraphics[width=0.48\linewidth]{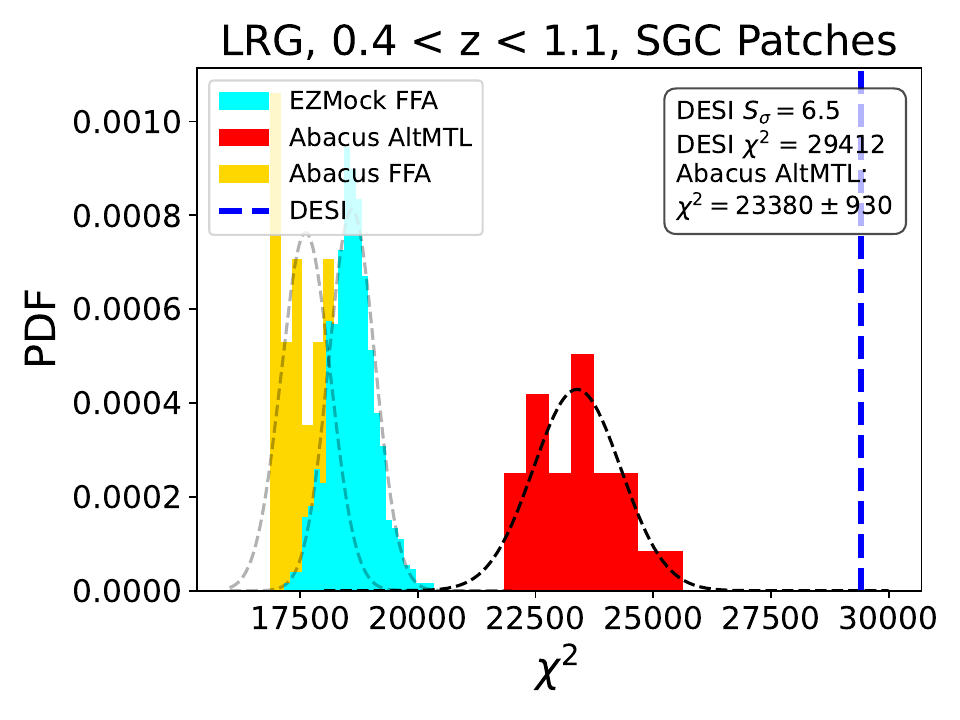}

    \caption{%
    While \cref{fig:auto-full-ngc-sgc} shows the auto-analysis on the full NGC and SGC, here we perform the auto analysis on the four NGC ($\num{5060} \times 4 = \num{20240}$ d.o.f.) and three SGC patches (\num{15180} d.o.f.) separately and then combine them, obtaining a similar significance (NGC: 10.5$\sigma$, SGC: 6.5$\sigma$, vs.\ 11.4$\sigma$ and 7.0$\sigma$ for the full). This figure shows that the discrepancy between the auto and cross analyses is not due to the differences in footprint. For discussion of the differences within each panel between the different fiber-assign methods and mock types, we refer back to \cref{fig:auto-full-ngc-sgc}'s caption.}
    \label{fig:auto_patches_ngc_sgc}
\end{figure*}

\textit{Combined Patch Auto}---We then show the auto significance for the combination of all the patches used for our cross analysis (still separated by hemisphere). We do this to assess whether the lost area when we go from full sample to patches is impacting significance notably. We find that it does not---the combined patch auto (\cref{fig:auto_patches_ngc_sgc}, NGC: 10.5$\sigma$, SGC: 6.5$\sigma$) is only slightly lower than the full sample auto (\cref{fig:auto-full-ngc-sgc}, NGC: 11.4$\sigma$, SGC: 7.0$\sigma$), as expected given the slightly reduced area.

In \cref{fig:auto_patches_ngc_sgc}, the mean $\chi^2$ of \textsc{Abacus} AltMTL is quite different from the number of degrees of freedom, in both NGC and SGC. This is because we use the same covariance matrix as in the full analysis from \cref{fig:auto-full-ngc-sgc}, but the combined volume of the patches is less than the total volume of the data.
In detail, the mean $\chi^2$ matches $(N_{\mathrm{dof}} / f_{\mathrm{cov}} )(V_{\mathrm{unique}} / V _{\mathrm{cutsky}})$ to within \SI{3}{\percent}, where the coverage fraction $f_{\mathrm{cov}}$ is defined in Eq.~B.10 of \cite{krol_parity}. $f_{\mathrm{cov}}$ is numerically equal to 0.55 (0.59) for NGC (SGC), similar to the ratio between the sum of the patch volumes and the total data volume, \SI{2.43}{\per\hHubble\cubed\Gpc\cubed} / \SI{4.33}{\per\hHubble\cubed\Gpc\cubed} for NGC and \SI{2.09}{\per\hHubble\cubed\Gpc\cubed} / \SI{3.32}{\per\hHubble\cubed\Gpc\cubed}. $V_\mathrm{eff}$ is defined for each patch using \cref{eq:variance_scaling_veff}, and the replication factor of 1.16 (1.13) in NGC (SGC) accounts for the mismatch between the mock and data variances. The replication factors differ from those used in the full analysis due to the different geometry of the union of the patches.

\textit{Regions Auto}---We next examine the auto on the regions, which are cut to $0.4< z < 0.8$, as opposed to the full which runs to $z < 1.1$. The $n(z)$ is flat to $z = 0.8$ and declining for $z > 0.8$ (\cref{fig:template}); this is a main motivation for the cut on the regions. Relative to the full sample, they are also restricted to only areas of high completeness. Thus, the regions help us determine if our evidence may be an artifact of either possibly mischaracterized change in $n(z)$ (which the randoms are chosen to match), some other survey issue that creeps in at high redshift, such as the selection function, and/or low completeness. 

In the NGC regions we find $2.3\sigma$ and $4.1\sigma$ signals, which add in quadrature to $4.7\sigma$. In the SGC region we find $6.3\sigma$. These are smaller than the full sample NGC and SGC significances of 11.4$\sigma$ and 7.0$\sigma$, as we expect given the smaller volume due to the smaller area and the redshift cut. Combining all regions (and including an appropriate replication correction for the \textsc{abacus} mocks as we do so), we find a $6.6\sigma$ signal. This is smaller than combining our full sample NGC and SGC (13.4$\sigma$), again as expected. Overall, the regions show that our evidence in the auto is not an artifact of possible issues that enter at high redshift.

\textit{Combined Cross}---We also show (\cref{fig:cross_combined}) the result of combining the main text NGC and SGC cross analyses. We do not include cross-hemisphere cross-correlations as the NGC and SGC have different best values of $\bar{n}$ for their covariances, meaning they cannot simultaneously be diagonalized and the NGC$\times$SGC is less effective. In the combined cross we find the data is consistent with the mocks, and thus, no evidence for PV. 

\textit{Covariance Calibration Choice}---We next explore our covariance matrix calibration, asking how much difference is made by the choice of mocks used to calibrate it. We calibrate the covariance using each of: \textsc{EZMock} FFA (main text), the \textsc{Abacus} AltMTL, and the \textsc{Abacus} FFA. We then analyze each mock using the covariance calibrated to that mock type. We also analyze the data using each covariance. This is shown in \cref{fig:covar-calib}. It is clear from inspection that the auto signal remains regardless of the choice of mocks.
%the choice we have made in the main text is most conservative, and that the other choices of calibration would lead to higher significances.

\textit{Covariance Half-Inverse Test}---In \cref{fig:covariance_test} we show the ``half-inverse'' test described in the main text. We also display a histogram of the matrix elements of this test and compare with a Gaussian; the test elements do not follow it perfectly. This suggests the deviations between the analytic covariance and the mock-based covariance are not entirely random and due to a finite number of mocks. This must be due to higher-order statistics, which our analytic covariance neglects, entering the true mock-based covariance. These could stem from dynamics, Halo Occupation Distribution (HOD) modeling, RSD, fiber assignment, or all of these; we do not speculate here on which, if any, is the dominant cause.

\textit{Variance Scaling}---\cite{krol_parity} derived a formula (their Eq.~B.6) for the variance of the null distribution for the auto $(\chi^2)$ as a function of an effective volume $V_{\mathrm{eff}}$ defined in their Eq.~B.1. The formula has one free parameter, a threshold volume $V_{\mathrm{thresh}}$ related to the scale at which non-Gaussian effects become important in the covariance. We have:
\begin{equation}
    \label{eq:variance_scaling}
    \frac{\mathrm{{Var}}(\chi^2)}{V_{\mathrm{fid}}^2} \approx \frac{2 N_{\mathrm{dof}}}{V_{\mathrm{eff}}^2} \left(1 + \frac{V_{\mathrm{thresh}}}{V_{\mathrm{eff}}} \right),
\end{equation}
with
\begin{equation}
    \label{eq:variance_scaling_veff}
    V_{\mathrm{eff}} \equiv \frac{V_{\mathrm{fid}}\, N_{\mathrm{dof}}}{\mathrm{Tr}\left(C_{\mathrm{ana}}^{-1} C_{\mathrm{mock}}\right)},
\end{equation}
with $V_{\mathrm{fid}}$ a fiducial volume (from \cref{tab:optimal_nbar_volume}). We obtain $V_{\mathrm{thresh}}$ by fitting the variances we obtain from the full sample, the patches, and the regions. We then show how well the variances match the relation (\cref{fig:variance-scaling}) for all mock types (we let $V_{\mathrm{thresh}}$ be different for each mock type). This is further evidence that we are handling the width of the null distributions correctly. In the lower panel of \cref{fig:variance-scaling}, we also show, for the 25 \textsc{abacus} mocks, the 1$\sigma$ errors, to give a sense of how well they fit the scaling relation. Save for two points, they are very consistent. We also compare the $\chi^2_\times$ distribution widths to the zero-free-parameter scaling for it as a function of $V_{\mathrm{eff}}$, Eq.~B.11 in \cite{krol_parity}. This matches $\sigma(\chi^2_\times)$ to within \SI{3}{\percent} for the NGC patches, the SGC patches, and the three NGC + SGC regions.

%\textbf{ZS: we need to make it clear we fit the covariance for each region separately, in the main text. I think we do that, anyway!}

\textit{Mock–data Consistency of 2PCF, 3PCF, and even 4PCF}---In \cref{fig:2_3_T2}, we examine the consistency of the 2PCF and 3PCF between mocks and data. In \cref{fig:even_4_T2}, we do so for the even-parity, connected 4PCF, see \textit{e.\,g.} \cite{even-4pcf, hou_even_desi, ortola_even_desi}. We do this by measuring these statistics on the mocks, and then using a mock-based covariance matrix to compute an overall significance for each. 

In detail, for \cref{fig:2_3_T2,fig:even_4_T2}, we exclude one \textsc{EZMock} at a time, use the other 999 \textsc{EZMocks} to compute a purely mock-based covariance, and then treat the excluded mock as if it is data and compute its $T^2$. For \textsc{Abacus} and for the DESI data, we do not need to perform this exclusion and use all \num{1000} \textsc{EZMocks} for the covariance. This procedure gives the distribution shown. We then compute a PTE for the data with respect to the mocks' distribution, and convert this PTE to the equivalent fluctuation (in number of $\sigma$) on a Gaussian distribution. The smallness of the equivalent number of $\sigma$ shows the good consistency of mocks with data. 

We show one (different) representative patch for each statistic, but we have performed this check for all patches and regions and the results are similar.
As mentioned in the main text, while the agreement between data and mock 2PCF, 3PCF, and even-parity 4PCF is necessary to ensure unbiased measurement of the parity-odd 4PCF variance, it is not sufficient, as discussed in the main text.

\begin{figure*}
    \centering
    \includegraphics[width=0.45\linewidth]{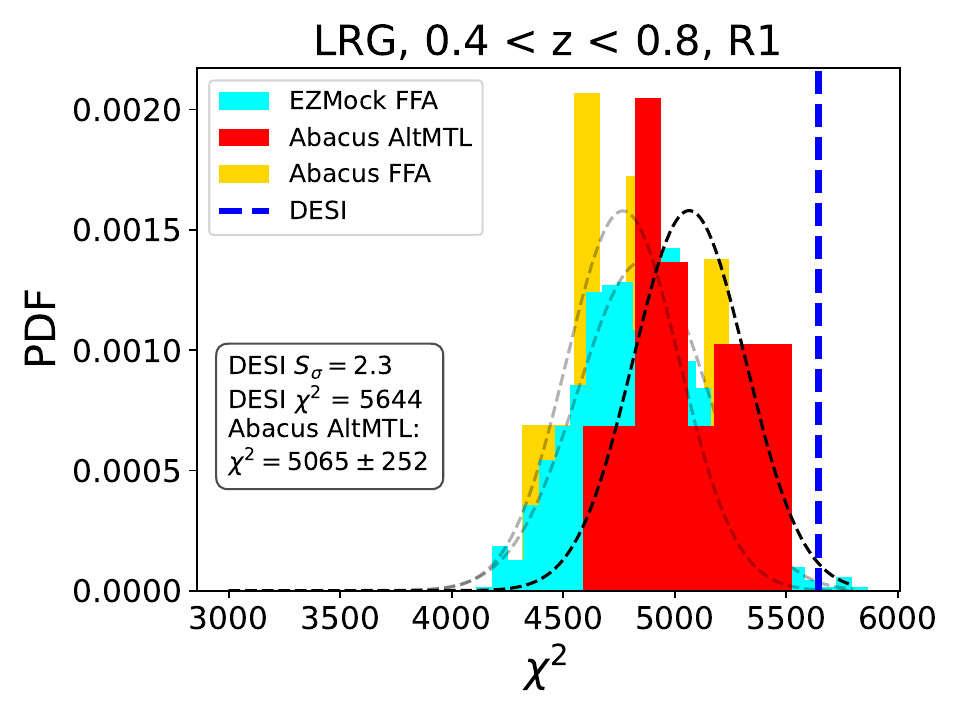}%
    \hfill%
    \includegraphics[width=0.45\linewidth]{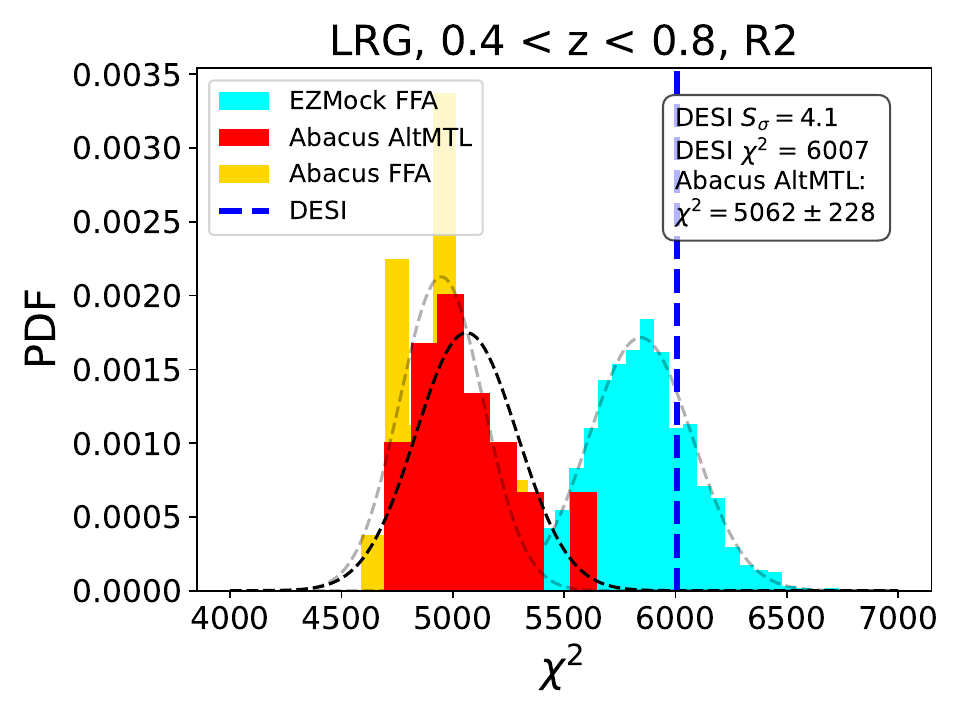}

    \includegraphics[width=0.45\linewidth]{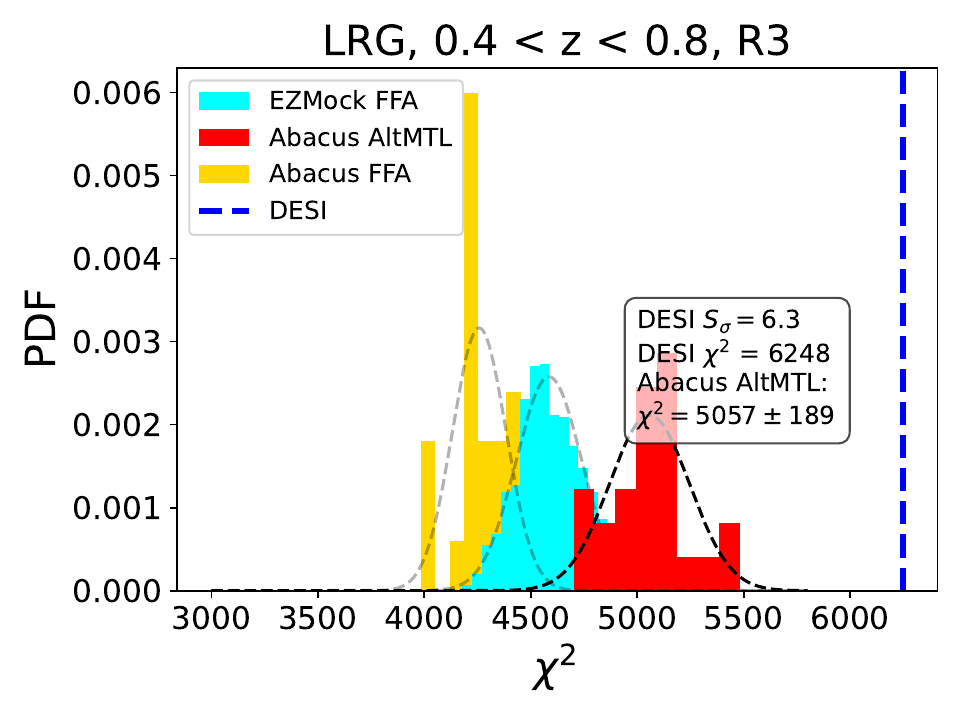}%
    \hfill%
    \includegraphics[width=0.45\linewidth]{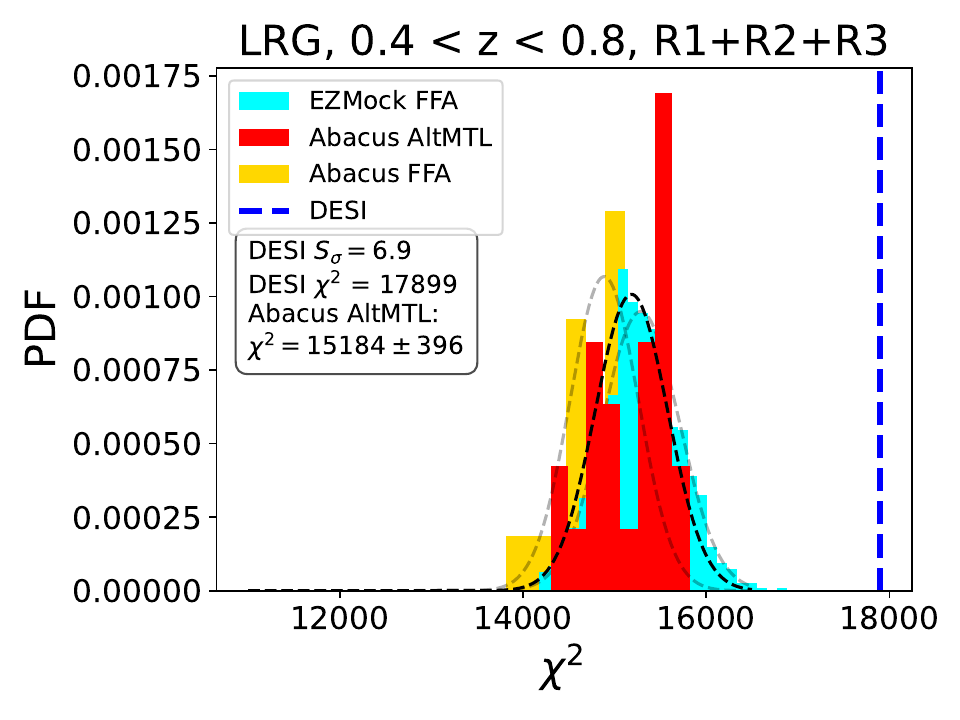}

    \caption{Auto analysis on each of the regions defined in \cref{fig:footprint} (R1 and R2 in the NGC, and R3 as the entire SGC), as well as their combination. We see $2.3\sigma$ evidence for PV in R1, $4.1\sigma$ in R2 (both NGC, \textit{upper row}), $6.3\sigma$ in R3 (SGC, \textit{lower left}) and a total significance of $6.9\sigma$ (\textit{lower right}). This final panel has $\num{5060} \times 3 = \num{15180}$ d.o.f. Unlike \cref{fig:auto_patches_ngc_sgc}, $V_{\mathrm{eff}}$ in the covariance used is normalized separately for each region, so that $\langle \chi^2 \rangle$ for \textsc{Abacus} AltMTL matches the number of degrees of freedom in each panel.}
    \label{fig:auto_regions_all_three}
\end{figure*}

%%%%Cross
\begin{figure}
    \centering
    \includegraphics[width=0.9\linewidth]{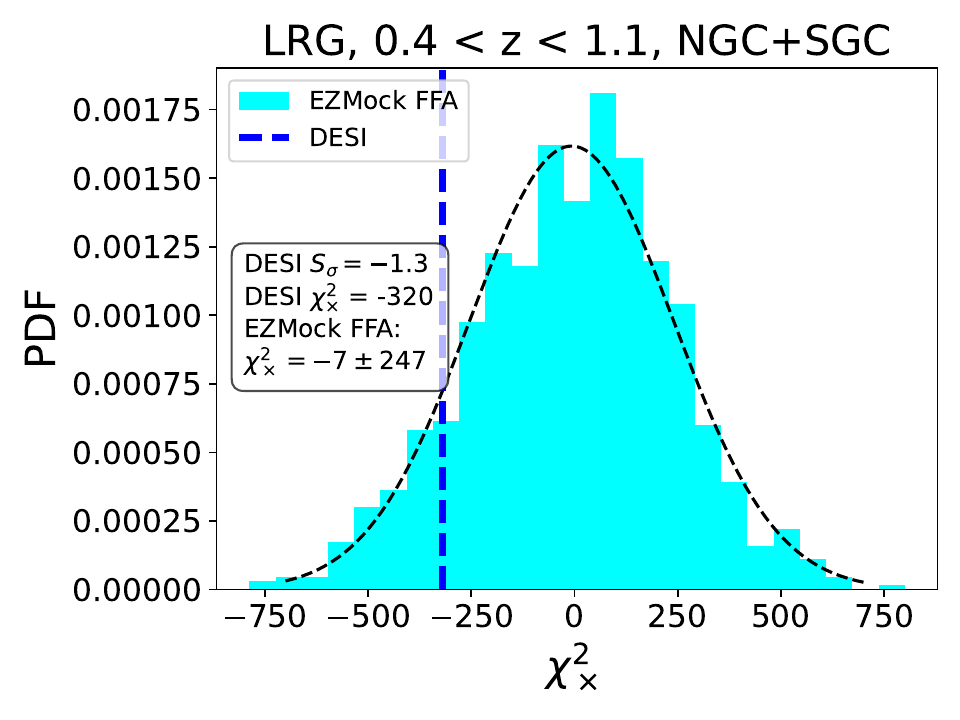}

    \caption{Combination of the NGC and SGC cross analyses in \cref{fig:combined_cross}, formed by summing $\chi^2_\times$ in both hemispheres. We do not form any cross-hemisphere cross-correlations, as explained in the main text.}
    \label{fig:cross_combined}
\end{figure}

\begin{figure*}
    \centering
    \includegraphics[width=0.45\linewidth]{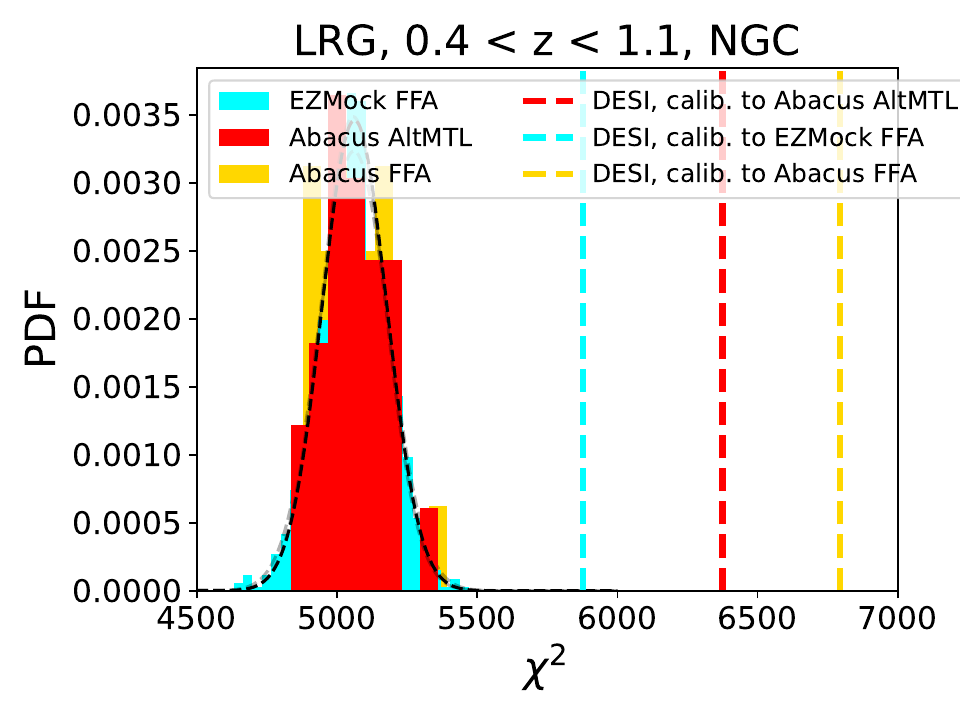}%
    \hfill%
    \includegraphics[width=0.45\linewidth]{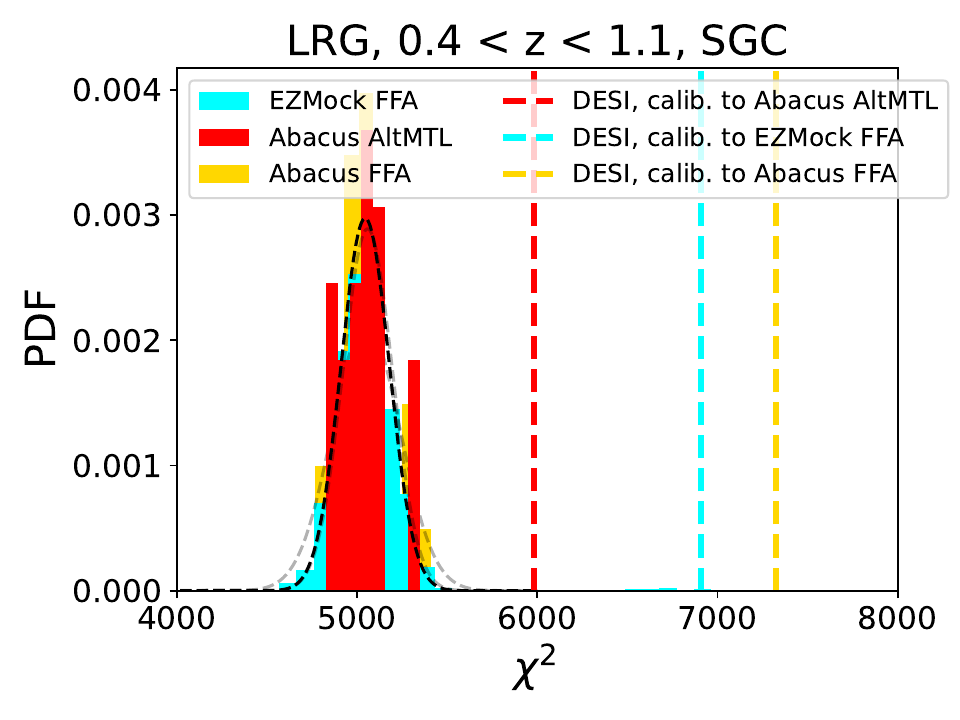}

    \caption{Here we show the results of calibrating our analytic covariance to each type of mock and analyzing the mocks and the data with that calibrated covariance; left is NGC and right is SGC. The DESI $\chi^2$ calculated using each calibrated covariance is shown by the dashed line, and each dashed line should be compared to the histogram of mock $\chi^2$ with the same color. The significance 
    in NGC is 6.6$\sigma$ (15.2$\sigma$)
    when calibrating to \textsc{EZMock} FFA (\textsc{Abacus} FFA), compared to 11.4$\sigma$ in \textsc{Abacus} AltMTL. In SGC, the significance is 9.9$\sigma$ (16.4$\sigma$) when calibrating to \textsc{EZMock} FFA (\textsc{Abacus} FFA), compared to 7.0$\sigma$ for \textsc{Abacus} AltMTL.}
    \label{fig:covar-calib}
\end{figure*}

\begin{figure*}
    \centering
    \includegraphics[width=0.48\linewidth]{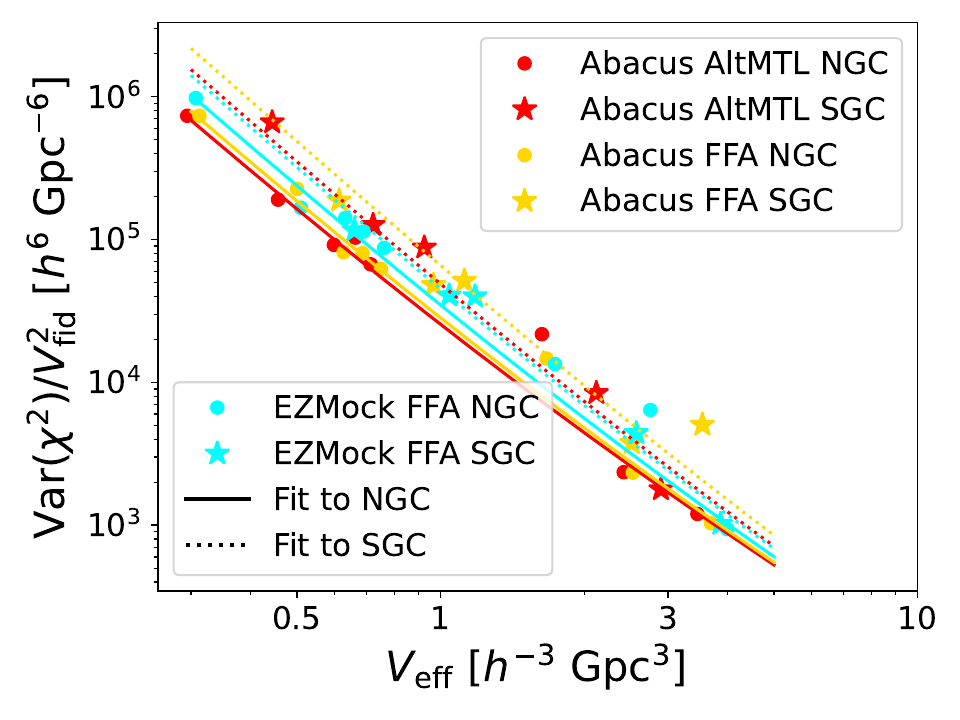}
    \includegraphics[width=0.48\linewidth]{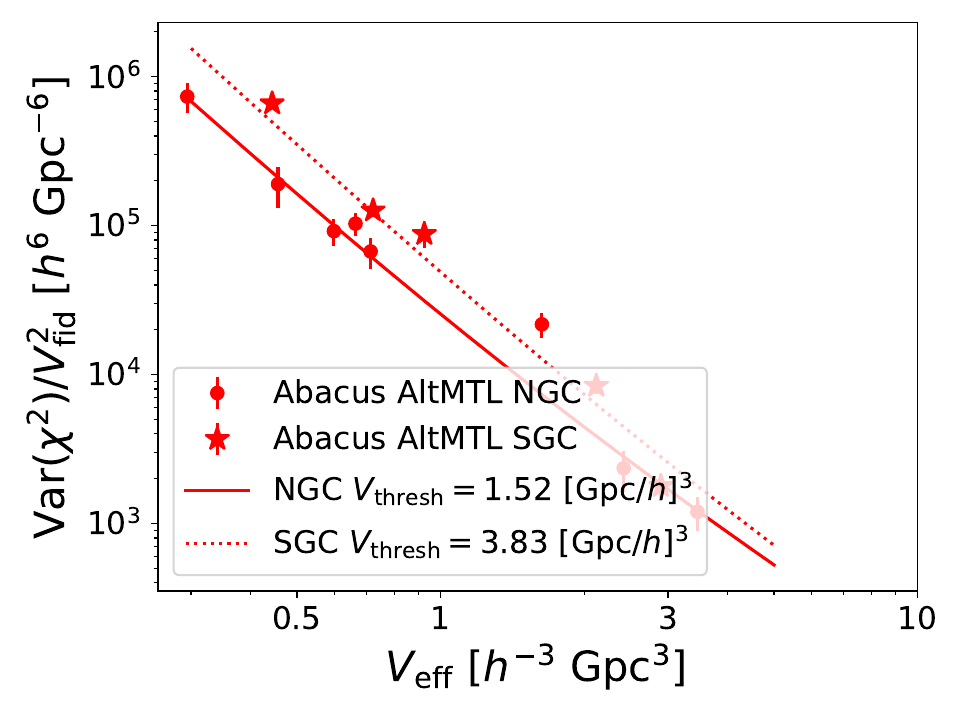}

    \caption{\textit{Left:} Variance of the auto $\chi^2$ as a function of effective volume; it is well described by the formula of \cite{krol_parity}, which we duplicate here (\pcref{eq:variance_scaling}). $V_{\textrm{fid}}$ is the fiducial volume of the covariance used to compute $\chi^2$, from \cref{tab:optimal_nbar_volume}, and $V_{\textrm{thresh}}$ is the best-fit from \cref{eq:variance_scaling}. Top panel shows \textsc{EZMock} FFA (cyan), \textsc{Abacus} AltMTL (red) and \textsc{Abacus} FFA (yellow) and their respective best-fits. \textit{Right:} We reproduce the \textsc{Abacus} AltMTL points and fits, but with error bars included from the standard error of the mean over 25 mocks. We do not show the \textsc{EZMocks} as their errors would be too small to be visible. Broadly, the mocks' variance is consistent with the formula of \cite{krol_parity} within the error bars, save for two points (one dot, one star) around $V_{\mathrm{eff}} \approx \SIrange{1.5}{2}{\per\hHubble\cubed\Gpc\cubed}$.
    %Also notably, for the NGC, $V_{\mathrm{thresh}} = \SI{1.52}{\per\hHubble\cubed\Gpc\cubed}$ is less than the volume of both one NGC region and the full NGC, meaning that in those the assumption of a GRF to compute the analytic covariance is independently justified.
    %\ak{I think this is kind of a weak argument (since Vthresh is not much less than Veff, the term isn't negligible; also what about SGC?). I'd prefer to remove this sentence.}
    }
    \label{fig:variance-scaling}
\end{figure*}

\begin{figure*}
    \centering
    \includegraphics[width=.48\linewidth]{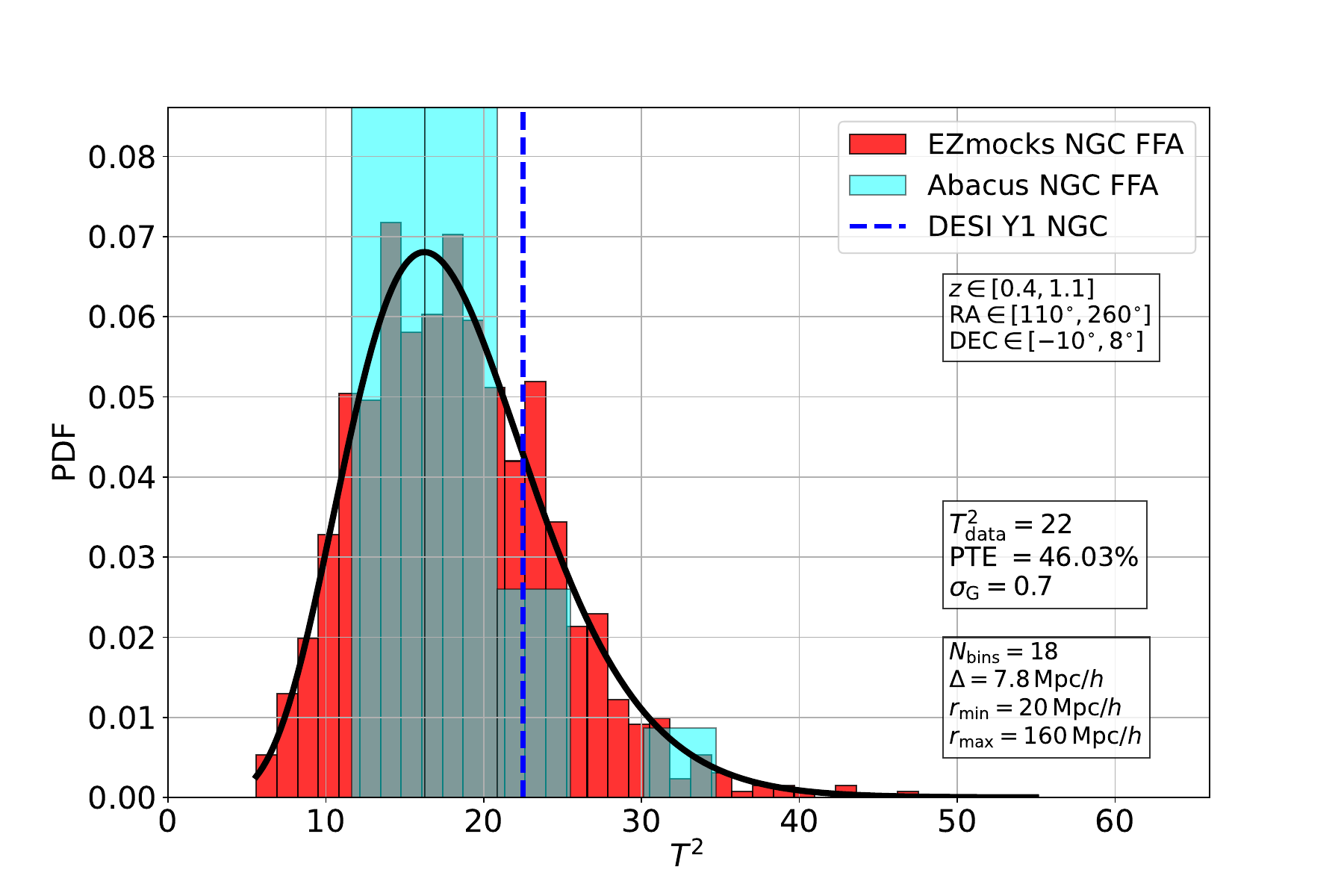}%
    \hfill%
    \includegraphics[width=.48\linewidth]{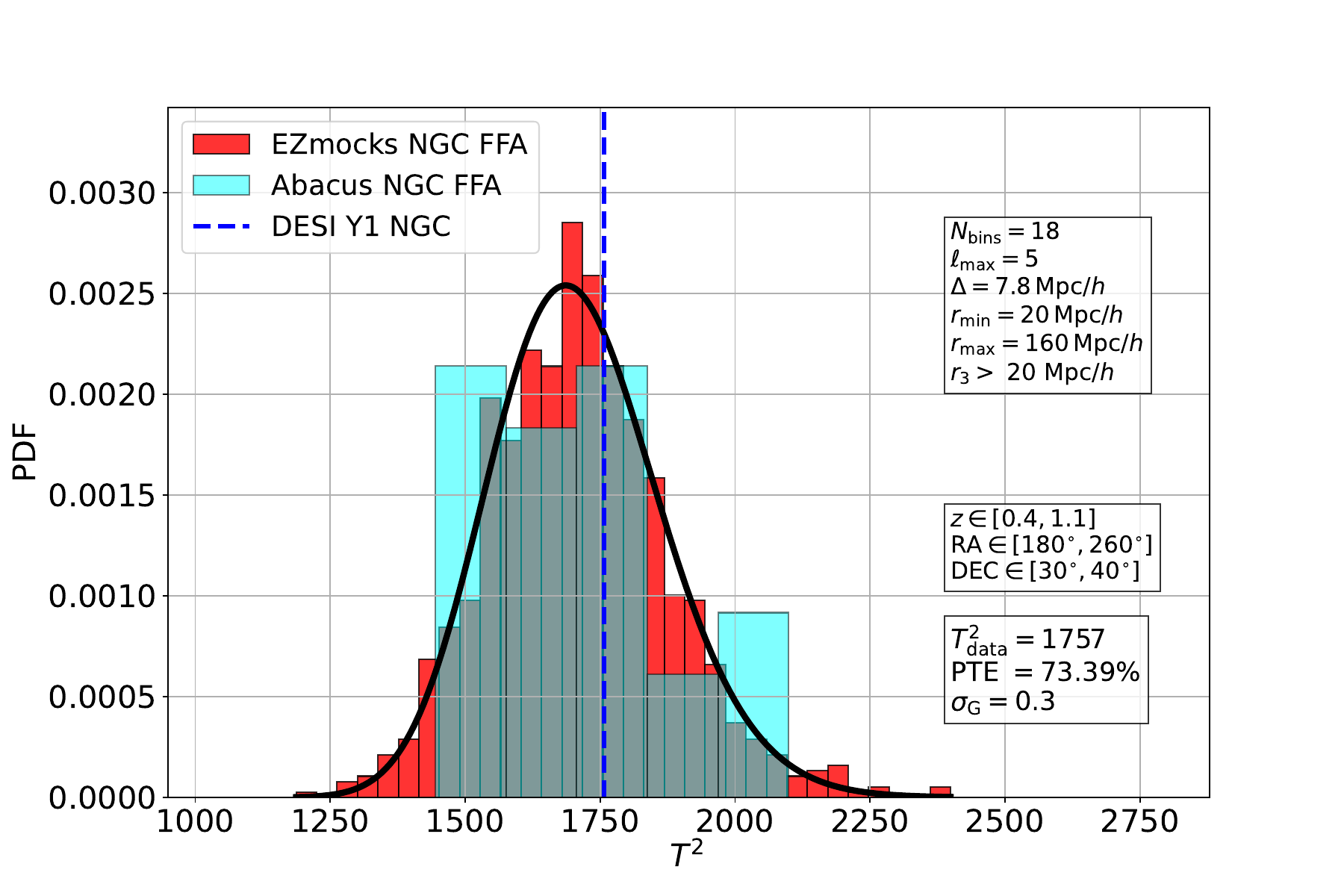}

    \caption{Distribution of $T^2$ values for the 2PCF and 3PCF, with the same minimum side length (\SI{20}{\per\hHubble\Mpc}) as used in our odd 4PCF analysis. Details of how this was produced are in the Appendix main text. \textit{Left:} For the 2PCF, the uppermost legend displays the patch, the middle text box the $T^2$ of the data, the associated PTE, and finally, the equivalent number of $\sigma$ to that PTE on a Gaussian distribution. Here we find the data is equivalent to a $0.7\sigma$ event on a Gaussian, meaning the data 2PCF is highly consistent with the mocks. The bottom legend displays the number of bins in pair separation, the bin spacing $\Delta$, and the minimum and maximum pair separations. \textit{Right:} For the 3PCF, the uppermost legend displays the number of bins, $\ell_{\mathrm{max}}$ used (including $\ell = 5$, for edge correction only), bin width $\Delta$, minimum and maximum triangle sides $r_1$ and $r_2$, and minimum $r_3$. The middle legend shows the patch, the bottom the $T^2$ of the data, the associated PTE, and the equivalent number of Gaussian $\sigma$, analogously to the left-hand panel. This last is $0.3\sigma$, so the data 3PCF is very consistent with the mocks.}
    \label{fig:2_3_T2}
\end{figure*}

\begin{figure*}
    \centering
    \includegraphics[width=\textwidth, height=0.9\textheight, keepaspectratio]{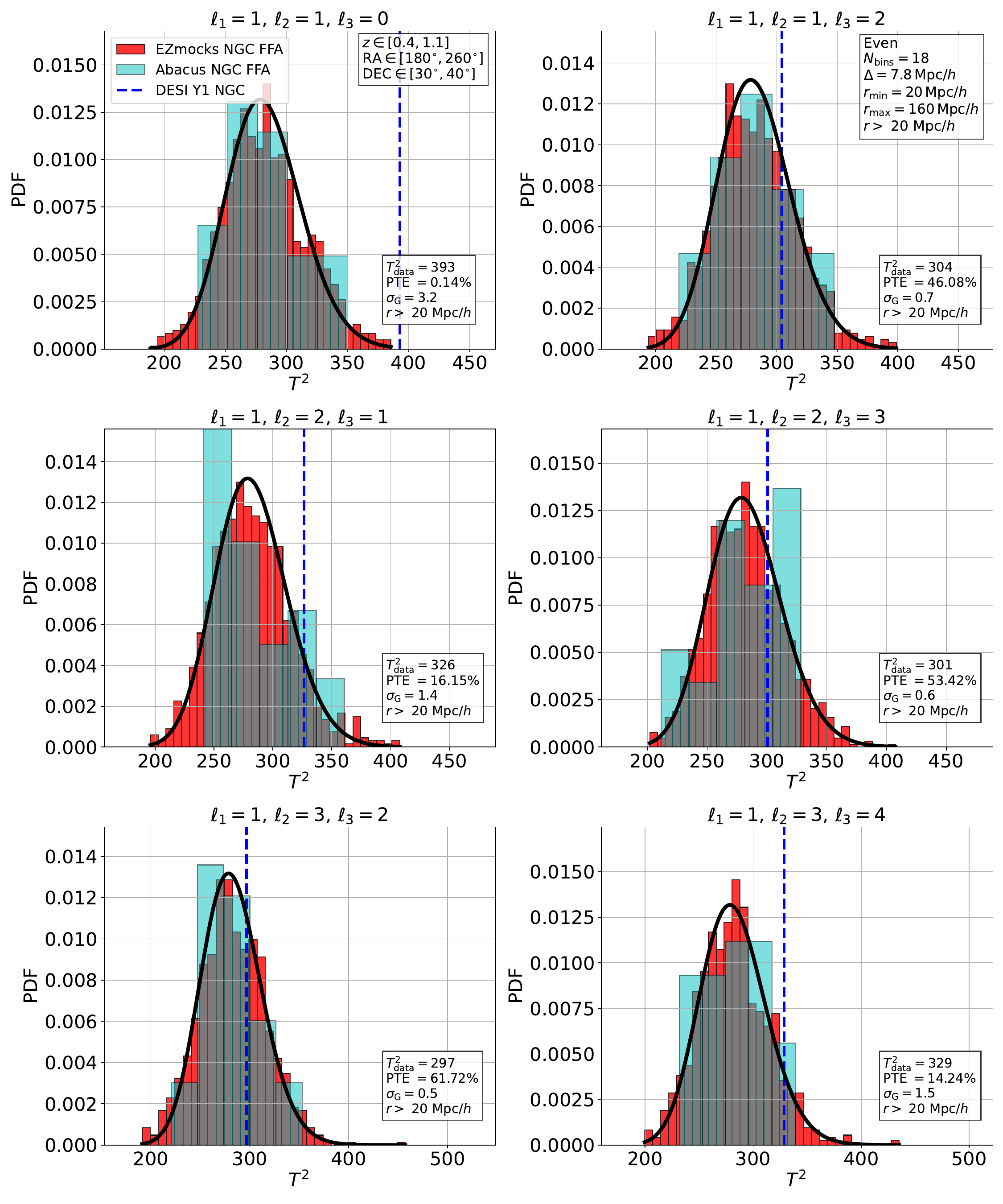}% Adjust scaling as needed

    \caption{Distribution of $T^2$ values for the even connected 4PCF, with the same minimum side length (\SI{20}{\per\hHubble\Mpc}) as used in our odd 4PCF analysis. Details of how this was produced are in the \hyperref[sec:appendix]{Appendix} main text, with further detail on reading the legend in \cref{fig:2_3_T2}'s caption. The key point is that $\sigma_{\mathrm{G}}$, the fluctuation (in sigma) to which the PTE is equivalent on a Gaussian distribution, is reasonable for each panel, showing the consistency of the even 4PCF for data and mocks. Here each panel is analyzed separately as we lack enough mocks to do a full mock-based covariance over all panels.}
    \label{fig:even_4_T2}
\end{figure*}

\begin{table*}
    \centering

    \begin{tabular}{l @{\quad} S[table-format=2.1] @{\quad} S[table-format=2.1]}
    \toprule
    \textbf{} & \textbf{\boldmath NGC ($\sigma$)} & \textbf{\boldmath SGC ($\sigma$)}
    \\
    \midrule
    Default & 11.4 & 7.0
    \\
    ``Most conservative width'' & 9.9 & 5.8
    \\
    ``Most conservative mean + width'' & 6.6 & 5.8
    \\
    Removing imaging weights & 8.9 & 6.2
    \\
    Removing redshift failure weights & 10.2 & 5.9
    \\
    % \hline
    % \textbf{} & \textbf{Optimal Volume \& $\overline{n}$ } & \textbf{Optimal Volume \& $\overline{n}$} & \textbf{} \\
    % \hline
    % \textsc{EZMock} & $6.436,\; 1.3 \times 10^{-4}$ & $3.838,\; 0.9 \times 10^{-4}$ \\
    % \textsc{Abacus} FFA & $4.26,\;.5 \times 10^{-4}$ & $3.539,\; 0.9 \times 10^{-4}$ \\
    \bottomrule
    \end{tabular}

    \caption{Statistical significance of the auto-analysis of NGC and SGC under different analysis choices and systematics tests. The top row gives the default analysis presented in the main text, comparing the DESI data to the \textsc{Abacus} AltMTL mocks. The second row compares the DESI data to the \textsc{Abacus} AltMTL mean, but divides by the largest standard deviation of: \textsc{Abacus} AltMTL, \textsc{Abacus} FFA, and \textsc{EZMock} FFA. The third row compares the DESI data to the largest mean and width of the mocks (the mean and width do not necessarily come from the same set). The next two rows display systematics tests in which we turn off the imaging systematics weights applied to the DESI data, and then the redshift failure weights. None of these changes make the auto evidence go away.}
    \label{tab:widths_zfail_imsys}
\end{table*}

\begin{table*}
    \centering

    \begin{tabular}{
        l @{\quad}%
        S[table-format=1.3] S[table-format=1.1e+1] S[table-format=1.2] @{\qquad}%
        S[table-format=1.3] S[table-format=1.2e+1] S[table-format=1.2]%
    }
    \toprule
    \textbf{} & \multicolumn{3}{c}{\textbf{NGC}} & \multicolumn{3}{c}{\textbf{SGC}}
    \\
    \textbf{} &
    {$V$ (\si{\per\hHubble\cubed\Gpc\cubed})} & {$\bar{n}$ (\si{\hHubble\cubed\per\Mpc\cubed})} & {$V_{\mathrm{mock}}/V_{\mathrm{unique}}$} &
    {$V$ (\si{\per\hHubble\cubed\Gpc\cubed})} & {$\bar{n}$ (\si{\hHubble\cubed\per\Mpc\cubed})} & {$V_{\mathrm{mock}}/V_{\mathrm{unique}}$}
    \\
    \midrule
    \textsc{Abacus} AltMTL & 4.319 & 1.6e-4 & 1.25 & 3.327 & 0.9e-4 & 1.15
    \\
    \textsc{EZMock} FFA & 6.436 & 1.3e-4 & {---} & 3.838 & 0.9e-4 & {---}
    \\
    \textsc{Abacus} FFA & 5.325 & 1.5e-4 & 1.25 & 4.070 & 0.9e-4 & 1.15
    \\
    % \hline
    % \textbf{} & \textbf{Optimal Volume \& $\overline{n}$ } & \textbf{Optimal Volume \& $\overline{n}$} & \textbf{} \\
    % \hline
    % \textsc{EZMock} & $6.436,\; 1.3 \times 10^{-4}$ & $3.838,\; 0.9 \times 10^{-4}$ \\
    % Abacus FFA & $4.26,\;.5 \times 10^{-4}$ & $3.539,\; 0.9 \times 10^{-4}$ \\
    \bottomrule
    \end{tabular}

    \caption{Optimal volume, number densities, and replication factor $V_{\mathrm{mock}}/V_{\mathrm{unique}}$ (only needed for \textsc{Abacus}) when the analytic covariance is calibrated to match each of the mock sets shown. We explored $\bar{n}$ from \SIrange{0.5e-4}{3e-4}{\hHubble\cubed\per\Mpc\cubed} in steps of \SI{0.1e-4}{\hHubble\cubed\per\Mpc\cubed}.
    \label{tab:optimal_nbar_volume}
    }
\end{table*}
%%%%ak{ZS: update volumes t match recent calculations! Also add rep factors}

\textit{Detailed Discussion of Systematics}---Here, we detail our systematics exploration. We first review what \cite{hou_parity_pub} explored to motivate the particular focuses we will then adopt.

\textit{Signal Level}---\cite{hou_parity_pub} found that it is very difficult for any systematic to produce a spurious parity-odd signal, by which we mean an odd 4PCF with non-zero average in the limit of an infinitely large survey. This is because the projection onto the isotropic basis destroys any part of the 4PCF that is not rotation-invariant, and to be parity-odd and rotation-invariant, a signal must be genuinely 3D, and cannot live in a 2D or 1D subspace \cite{iso_basis_pub, cahn_short_pub}. 

Plane-of-sky (2D) systematics \cite{hou_parity_pub} explored include variation of $n(z)$ with imaging depth (their App.~D), fiber collisions (their \S6.1.2), and tidal alignments (their \S6.2.2). Line-of-sight (1D) systematics explored included errors in the selection function (their \S6.1.1, with a toy model, and empirically-imposed radial distortions in their \S6.1.2), magnification bias (their \S6.1.3), and redshift-space distortions (their \S6.2.1). As the above makes clear, \textit{almost} all systematics live either in the plane of the sky (2D) or along the line of sight (1D). Regarding other systematics that are not simply 1D or 2D, \cite{hou_parity_pub} also quantified the impact of change in the fiducial cosmology (their \S6.3.1) and of survey geometry and the edge-correction procedure (their \S6.3.2), finding they are not significant effects.

One interesting exception to the 1D/2D separability that \cite{hou_parity_pub} found was an $x$-$y$ dependent, and asymmetric, redshift failure rate due to galaxies on opposite edges of the BOSS plates falling on two different cameras' edges, and where the cameras had slightly different efficiencies. This effect was also more pronounced for fainter galaxies, which \textit{ceteris paribus} tend to be at higher redshift, bringing in (Cartesian) $z$-dependence---hence, a truly 3D effect. 

\cite{hou_parity_pub} empirically doubled the rate of this effect's occurrence in the data, and showed it was not sufficient to shift their result substantially enough to explain away the statistical evidence. We would not necessarily expect this effect in DESI because the fibers' mapping to the spectrograph and cameras is different from BOSS's.
We also find that our results are not much altered if we turn off the redshift failure weights, which correct for the variation in LRG success rate with spectroscopic observing conditions.

\textit{Variance Level}---There are many systematics that, while they cannot produce a spurious signal, can increase the \textit{variance} of the data beyond what the covariance matrix models. Since the significance is computed using the inverse covariance, having the model covariance underestimate the true covariance in this way could produce a spuriously high result \cite{cahn_short_pub}. This is a serious concern and was studied extensively in \cite{hou_parity_pub}. Many of the possible issues in DESI would be the same as those in BOSS and thus were exhaustively explored in \cite{hou_parity_pub} \S6.

\textit{Fiber Assignment}---Thus, we focus here on the primary difference between DESI and BOSS---DESI uses \num{5000} robotic fiber positioners, each fixed in place and with some limited range of movement. Thus DESI has a more complicated pattern of missing galaxies than does BOSS (\textit{e.\,g.} \cite{Hand2018, Burden:2016cba, Hahn16, Pinol:2016opt, Pinon25} explore this in the 2-point statistics). This issue is here referred to as ``fiber assignment''. Since each galaxy must be assigned a fiber to have a spectrum taken, fiber assignment (or lack thereof) is what produces possible increased variance. As fiber assignment is essentially a random loss of galaxies in high-density regions, it can increase the variance, but does not produce a spurious ``infinite-volume-average’’ signal.

Fiber assignment only matters if the number of targets falling on a given part of the DESI plate during a given exposure exceeds the number of available fibers in that part. This is most likely to happen in the regions of highest projected density, which are also most likely to be regions of high 3D density.

\textit{Two Methods of Fiber Assignment}---There are two methods of fiber assignment used on the mocks. The first, AltMTL \cite{Lasker25}, is very close to what is done on the real data. It performs a complicated assessment of what has been done up to a given point in the survey---how complete each target class is, \textit{e.\,g.} LRGs, Emission Line Galaxies, quasars, etc., what the observing conditions on a given night are, and how likely each spectrum is to be obtained successfully. It is thus computationally expensive to apply to mocks.

The second method, FFA \cite{KP3s11-Sikandar}, is a much cheaper procedure tuned to reproduce the results of AltMTL. It has been found to give consistent results with AltMTL at the 2PCF level; in this work, we explore this 4PCF and summarize the 3PCF impact (more fully described in \cite{farshad_3pcf_bias, farshad_3pcf_bao}).

Since AltMTL is costly, it has been applied only to 25 \textsc{Abacus} mocks, and two \textsc{EZMocks}. The cheaper FFA has been applied to all \num{1000} \textsc{EZMocks} (including the two mentioned above) and also to the 25 \textsc{Abacus} mocks. Thus, we can make several comparisons.

First, we assess the impact of FFA vs. AltMTL on \textsc{EZMocks}. We cross-correlated the measured 4PCF of matched sets made by applying the AltMTL and FFA to the same \textsc{EZMock} and found high $r$-values, of order \SI{99}{\percent}, so this test was not a very powerful probe of fiber assign. 

Second, we assess the impact on \textsc{Abacus} mocks. Why does this matter? The \textsc{Abacus} mocks are fully $N$-body; hence we might hope they do a slightly better job than the \textsc{EZMocks} in capturing the small-scale clustering of galaxies, as well as perhaps having higher fidelity to the truth in the most over-dense, most evolved regions that are likeliest to be impacted by fiber assignment. 

This last comment requires a caveat, that in both types of mock, a HOD model \cite{Yuan22} is used to assign galaxies to halos, and this prescription is tuned to reproduce the observed clustering. Hence, it is not inarguable that \textsc{Abacus} does better than \textsc{EZMocks} with regard to the true clustering. It may therefore be preferable to frame this comparison as simply, there are two different methods for producing mocks---is our analysis robust against this choice?

Since FFA underestimates the parity-odd 4PCF variance compared to (the fully realistic) AltMTL, this could affect $T^2$ computed using the \textsc{EZMock} covariance in our compressed analysis (as in the even-parity analysis of \cite{hou_even_desi}). However, if we scaled the covariance matrix to account for the under-estimated variance, this would change $T^2$ of the data and the \textsc{AltMTL} by the same factor, and thus leave our significance unaffected. On the other hand, re-scaling the \textsc{EZMock} FFA variance would change the significance of the data relative to \textsc{EZMock}. In particular, if we scale the \textsc{EZMock} $\chi^2$ by the ratio of the \textsc{Abacus} AltMTL and FFA mean $\chi^2$ (equivalent to the fiber-assignment correction factor of \cite{hou_even_desi}, we would reduce the significance from 6.6$\sigma$ to 4.5$\sigma$ in NGC, and 5.7$\sigma$ to 3.2$\sigma$ in SGC.

\textit{Redshift Failures and Imaging Systematics}---We also explore redshift failure, since it is relatively straightforward to do so. We switch off the ``zfail'' weights on the data and recompute the significance of PV. We do the same for the imaging systematics (``imsys'') weights. These results are in \cref{tab:widths_zfail_imsys}. Neither changes our significance sufficiently to suggest redshift failure or imaging systematics as an alternative explanation of our statistical evidence for PV. 

\textit{Scale cuts}---Finally, we study the impact of changing the scale cuts used in \cref{fig:scale_cut}. We test imposing either a small-scale or large-scale cut, in both cases reducing the number of degrees of freedom roughly by half. We find that the significance is reduced by $\sqrt{2}$ in both cases, consistent with the signal preferring neither large nor small scales. 

%We note that the lack of scale-dependence in the signal may be evidence for a mock–data mismatch driving the auto $\chi^2$ results.

%%%%Complete vs Altmtl vs FFA
\begin{figure*}
    \centering
    \includegraphics[width=0.48\linewidth]{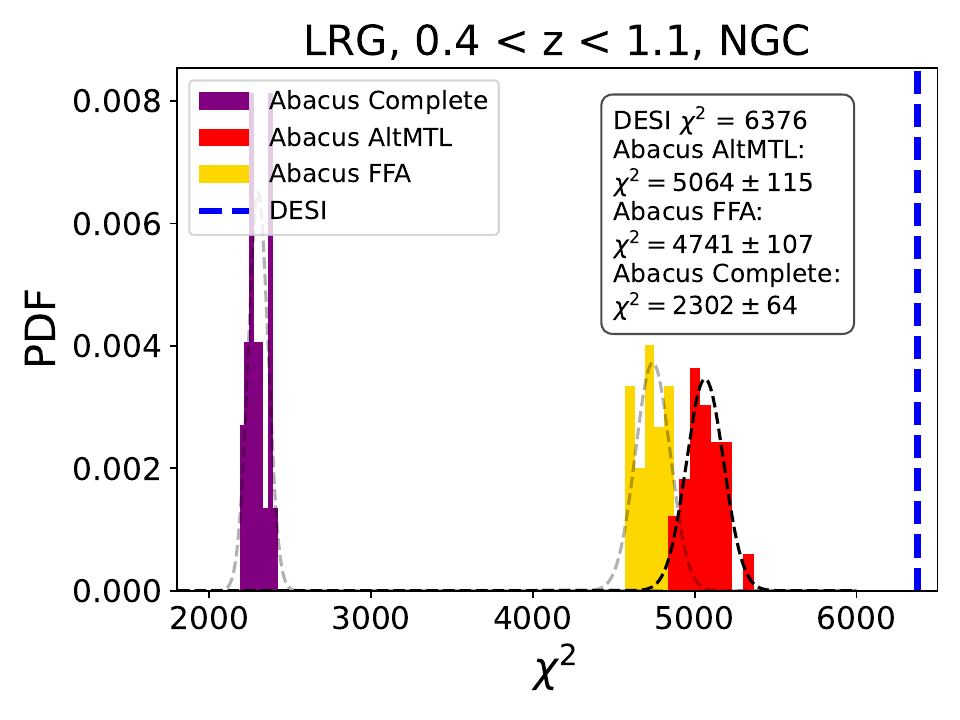}%
    \hfill%
    \includegraphics[width=0.48\linewidth]{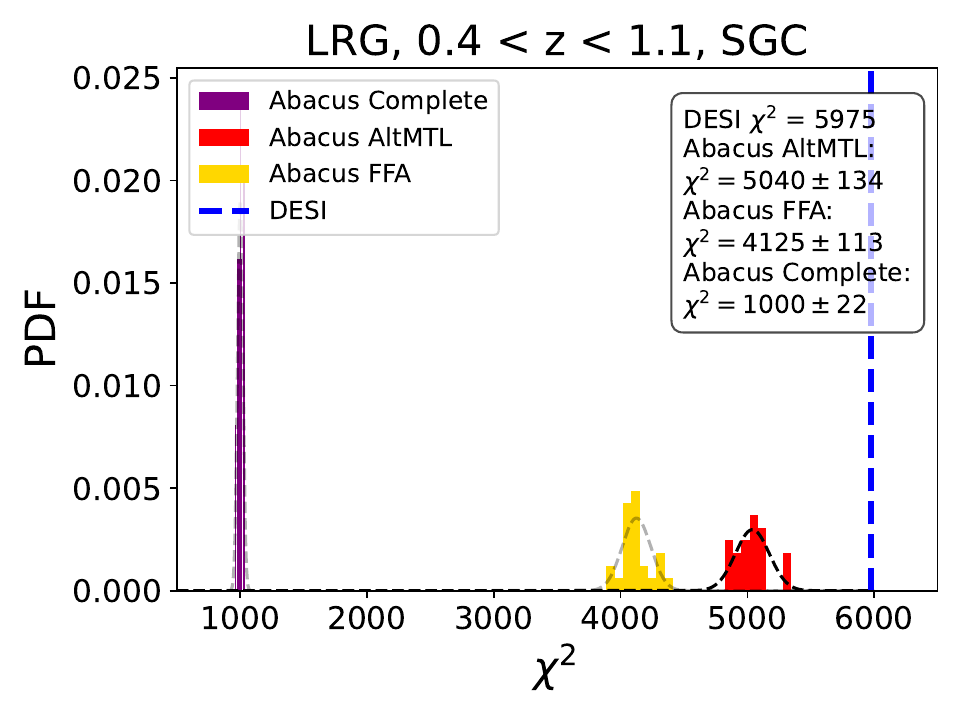}

    \caption{Here we show \textsc{Abacus} mocks for the full sample (\textit{left}: NGC, \textit{right}: SGC) with no fiber-assignment (``complete'', purple), with AltMTL (red) and with FFA (yellow), as well as the DESI data (dashed blue). Running the mocks through DESI fiber-assignment considerably changes $\chi^2$, and there is a small difference between the less realistic FFA and the more realistic AltMTL. The legend gives the center and width of each distribution.
    }
    \label{fig:complete_vs_altmlt_vs_FFA}
\end{figure*}

\begin{figure*}
    \centering
    \includegraphics[width=0.48\linewidth]{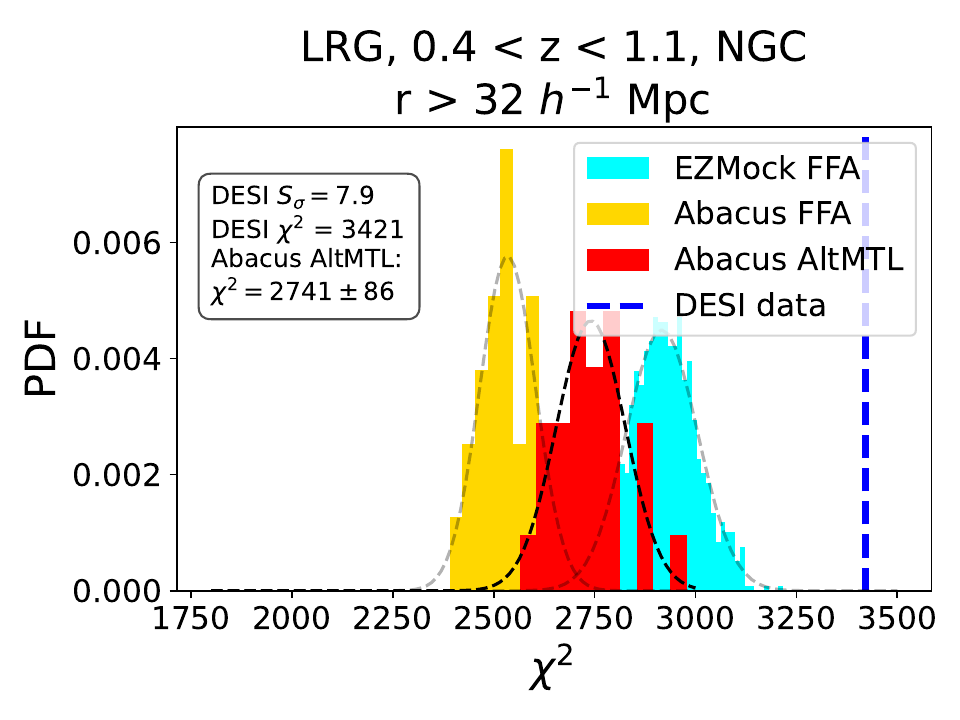}%
    \hfill%
    \includegraphics[width=0.48\linewidth]{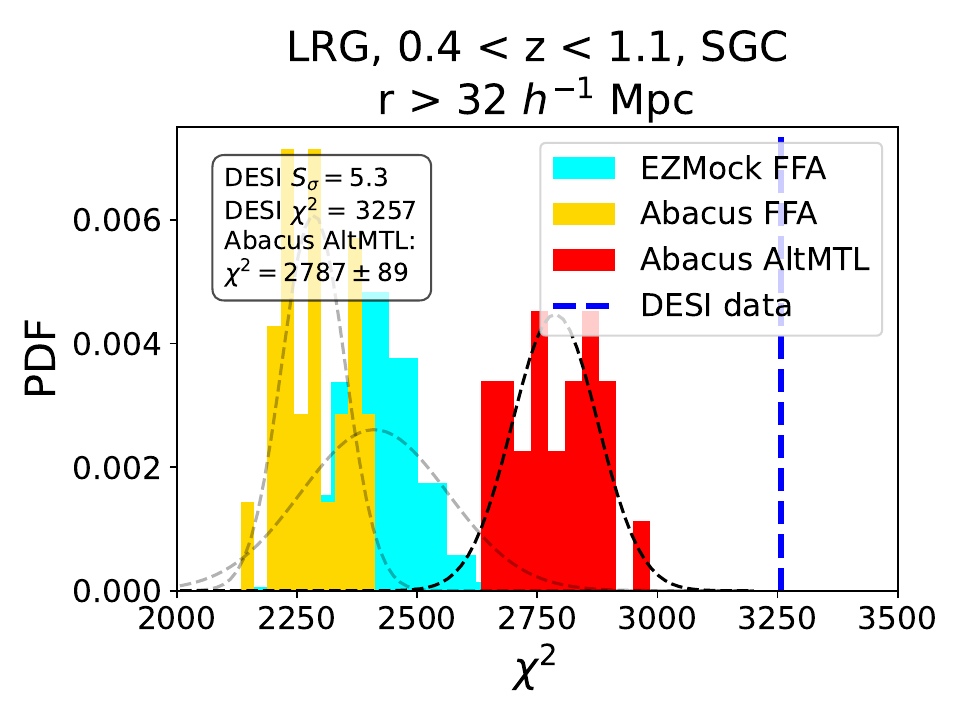}

    \includegraphics[width=0.48\linewidth]{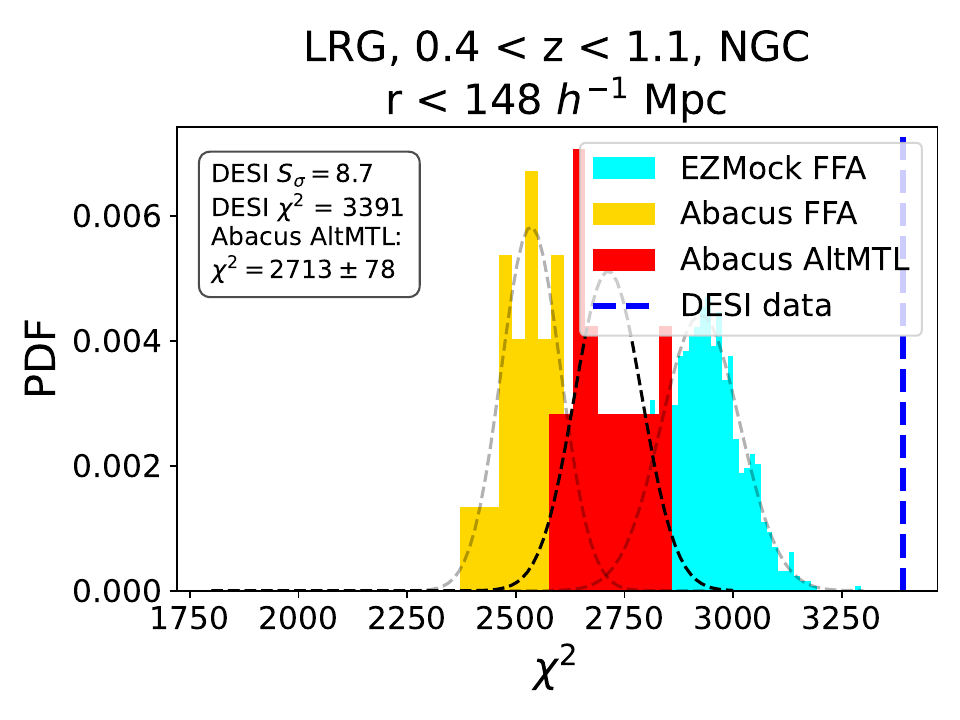}%
    \hfill%
    \includegraphics[width=0.48\linewidth]{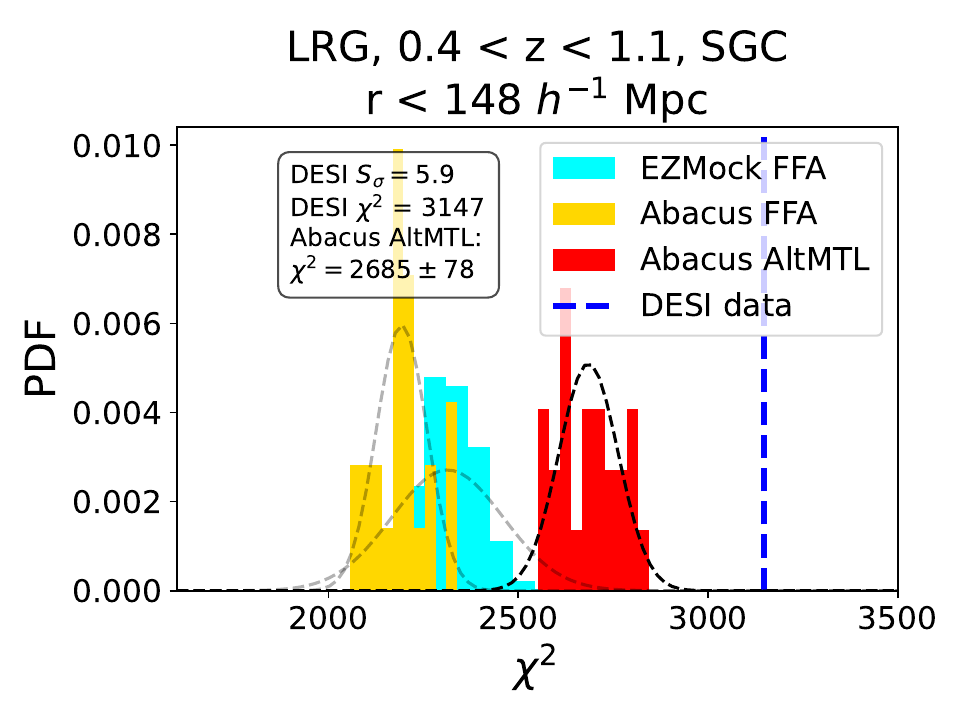}

    \caption{Significances in NGC (left) and SGC (right), when imposing additional scale cuts: either requiring all tetrahedron sides to be larger than \SI{32}{\per\hHubble\Mpc} (top), or all tetrahedron sides to be smaller than \SI{148}{\per\hHubble\Mpc} (bottom). In both cases, this reduces the number of degrees of freedom by roughly a factor of two, to \num{2760}, and the significance is consequently reduced by about \SI{40}{\percent}. Removing small scales is driven by the fact that the covariance and mocks may be less reliable on more non-linear scales; removing large scales by the fact that one can have imaging systematics that cause large-scale variations in the survey.}
    \label{fig:scale_cut}
\end{figure*}

%%%%Half-inverse
\begin{figure*}
    \centering
    %Alex include graphics as half inverse test of 111, histogram of 111, diagonal f 111, then histogrma of full matrix. This is the order I wrote the caption in.
    \includegraphics[width=0.48\linewidth]{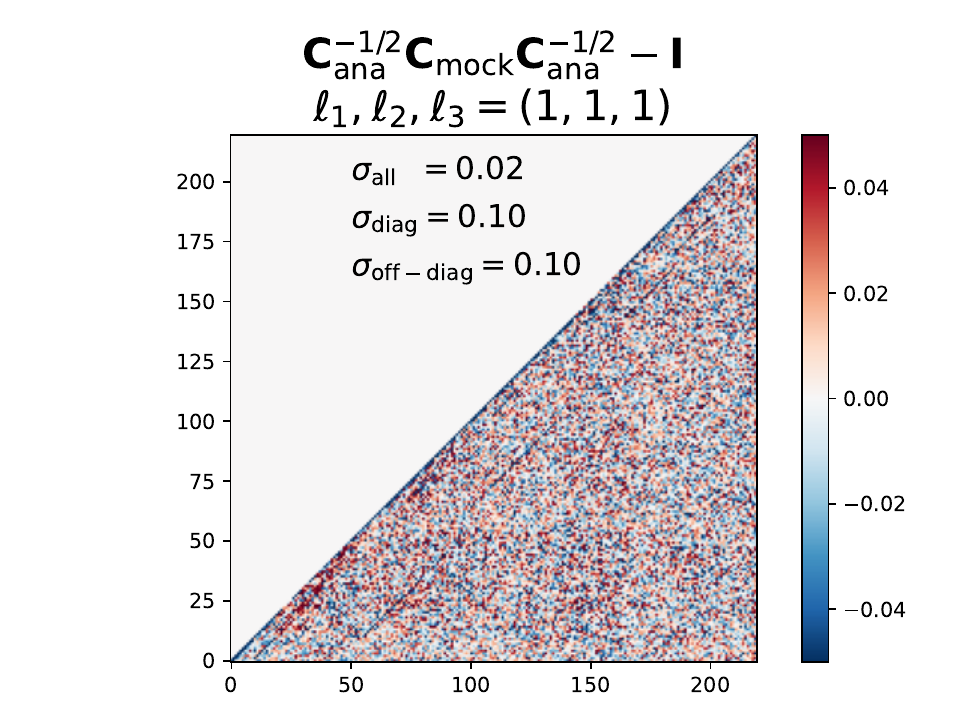}
    \includegraphics[width=0.48\linewidth]{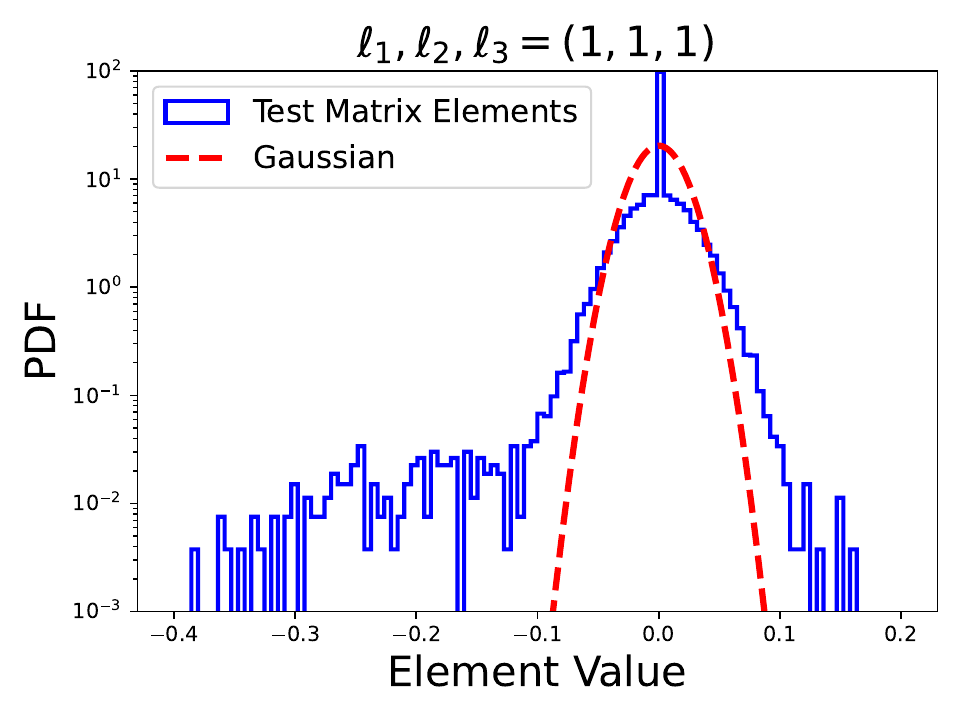}
    \includegraphics[width=0.48\linewidth]{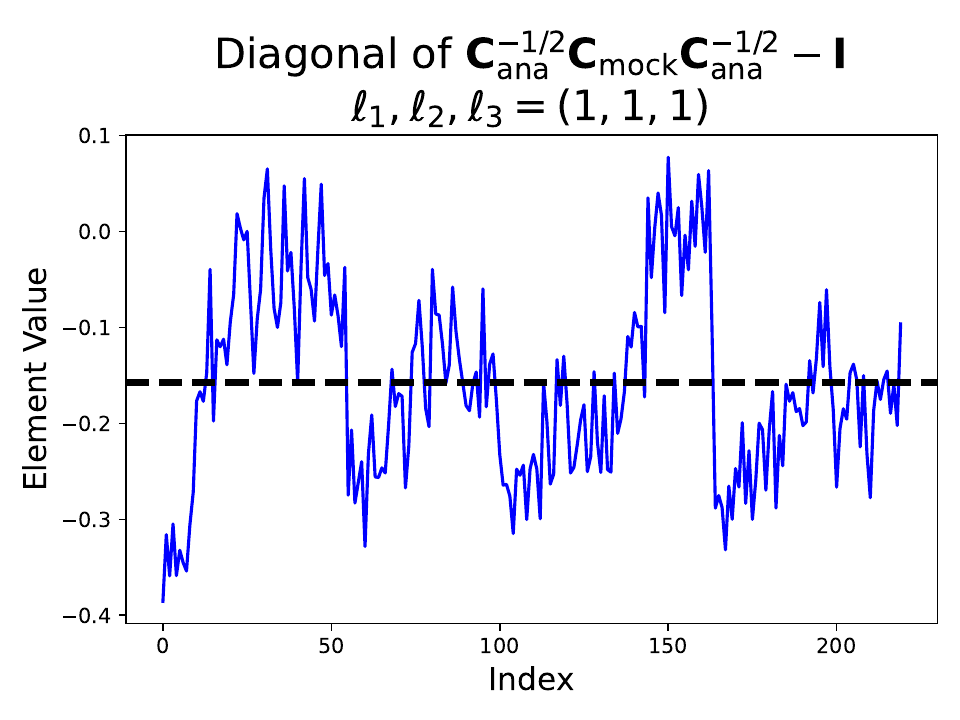}
    \includegraphics[width=0.48\linewidth]{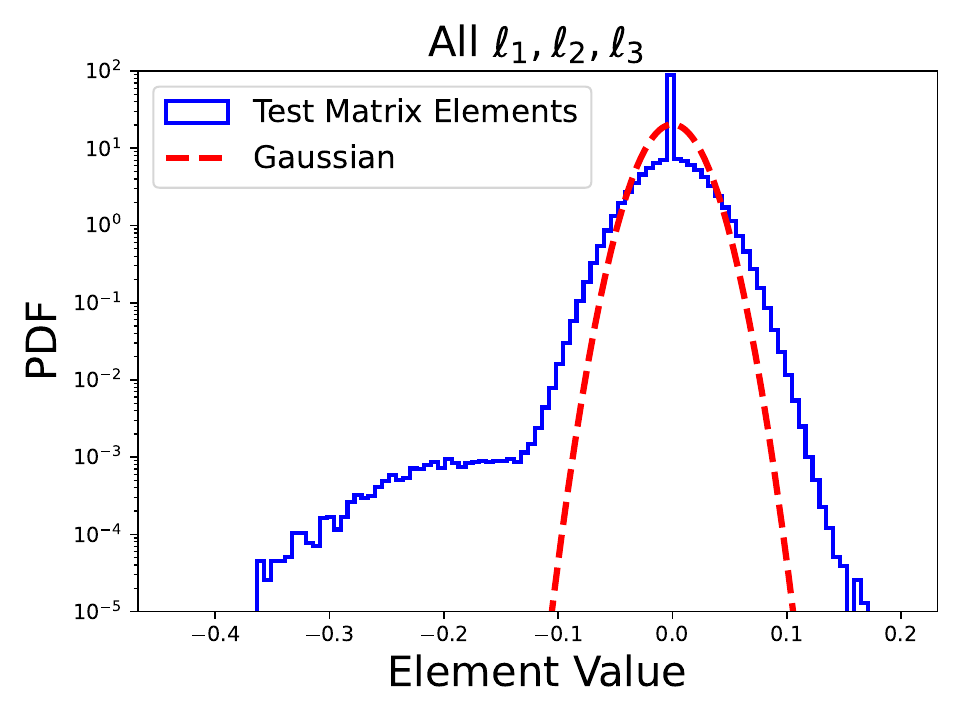}

    \caption{\textit{Top left:} %\ak{Make blue and red lines thicker}
    Test matrix formed by multiplying the inverse of the theory covariance by the EZMock-based (noisy) covariance and subtracting the identity matrix. If the inverse perfectly reflected the true (mock-based) covariance, this plot would be consistent with Gaussian noise, with $\sigma_{\mathrm{all}} = 1/\sqrt{N_{\mathrm{mocks}}}$. It is a known result that one should also have $\sigma_{\mathrm{diag}} = 2 \sigma_{\mathrm{off-diag}}$. For legibility we show only the $111$ mode, as the full matrix is $\num{5060} \times \num{5060}$ and is hard to read when plotted. We have inspected each mode's test matrix and confirmed the $111$ behavior is representative. \textit{Top right:} histogram of these test matrix elements compared with a Gaussian. We see that the distributions are comparable towards the center but the tails are much wider in our case than for the Gaussian; this is further discussed in the \hyperref[sec:appendix]{Appendix} main text. \textit{Bottom left:} diagonal of the test matrix and its average---we see there is an offset from zero. \textit{Bottom right:} histogram of the full test matrix for all modes. Overall, these panels show that our theoretical covariance is a reasonable but not perfect inverse for the noisy mock-based covariance. The negative tail suggests our analytic covariance \textit{over-estimates} the variance.\footnote{If we have $\mathbf{C}^{-1}_{\mathrm{ana}} \mathbf{C}_{\mathrm{mocks}} - \mathbf{I} = -\mathbf{\epsilon},\; \epsilon > 0$, then $\mathbf{C}_{\mathrm{mocks}} = \mathbf{C}_{\mathrm{ana}} ( \mathbf{I} - \mathbf{\epsilon})$, showing that the mock covariance would be typically less than the analytic in this case, meaning the latter \textit{over-estimates} the variance.}} 
    \label{fig:covariance_test}
\end{figure*}

%%%%Input power spectrum
\begin{figure*}
    \centering
    \includegraphics[width=0.48\linewidth]{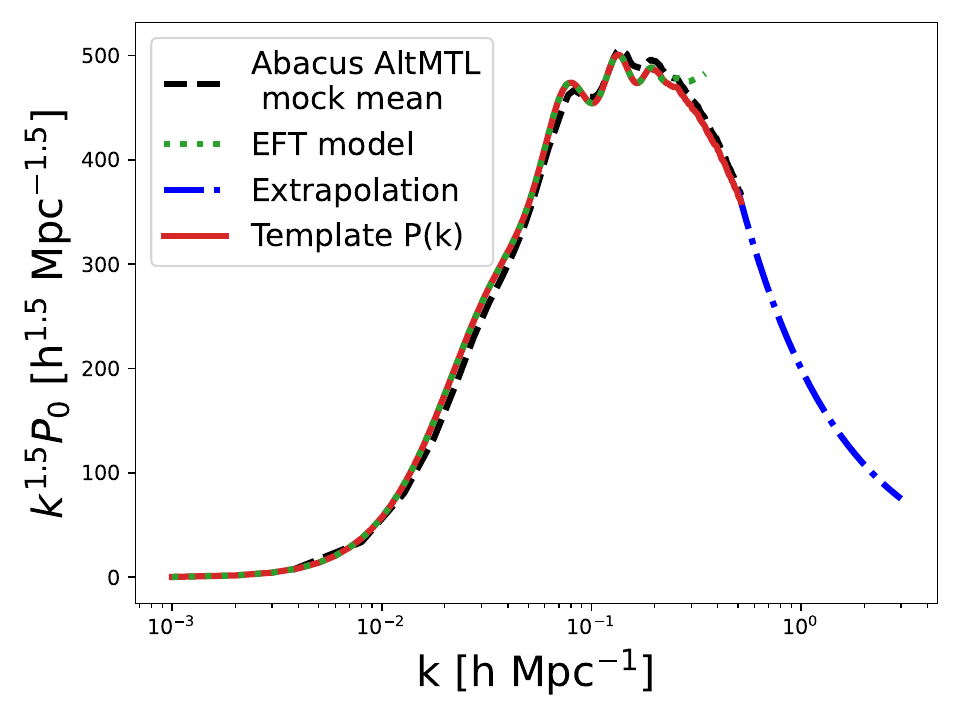}%
    \hfill%
    \includegraphics[width=0.48\linewidth]{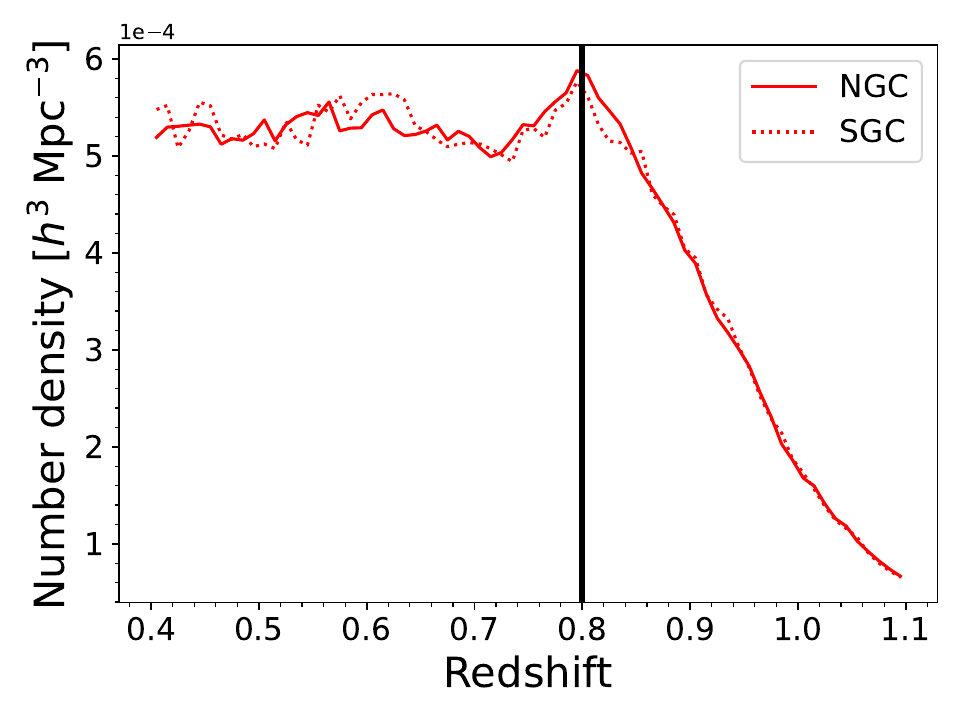}

    \caption{\textit{Left:} Power spectrum (red) for the analytic covariance. For $k < \SI{0.2}{\hHubble\per\Mpc}$, we use the best-fit EFT model (green); for $\SI{0.2}{\hHubble\per\Mpc} < k < \SI{0.5}{\hHubble\per\Mpc}$, the mean of the \textsc{Abacus} AltMTL mocks (black); and for $k > \SI{0.5}{\hHubble\per\Mpc}$, a power-law extrapolation (blue). We use the mean of the mocks where we do because this is outside the fitting range of the EFT model, and the counter-terms can cause the model to quickly diverge (\textit{i.\,e.} the green curve for $k > \SI{0.2}{\hHubble\per\Mpc}$). \textit{Right:} $n(z)$ for the full LRG samples (NGC solid, SGC dotted). The redshift cut for the \textit{regions} defined in \hyperref[sec:data_cuts_mocks]{Data, Cuts, Mocks}, $z=0.8$, is chosen because $n(z)$ begins its decline there.}
    \label{fig:template}
\end{figure*}

\clearpage

\section*{Author Affiliations}

\printaffiliations

\section*{Acknowledgments} 
ZS \& AG acknowledge funding from NASA grant number 80NSSC24M0021; ZS acknowledges funding from UF Research AI award \#00133699. AK was supported as a CITA National Fellow by the Natural Sciences and Engineering Research Council of Canada (NSERC), funding reference \#DIS-2022-568580. All UF authors and AK and SM gratefully thank UF Research Computing, especially Ying Zhang and Alex Moskalenko, and Kaleb Smith of NVIDIA's UF Research Center. We also especially thank the DESI Publications Board handlers, Jiaxi Yu and Alejandro Aviles, the internal DESI referees, Zachery Brown and Lado Samushia, as well as Robert Cahn and Jiamin Hou for insightful comments and sage counsel. We thank Jiamin Hou also for initial preparation of the data used as well as substantial contributions to the NPCF infastructure enabling this analysis. We thank all Slepian group members for useful conversations, as well as Tamara Davis, Kyle Dawson, Daniel Eisenstein, Douglas Finkbeiner, Hector Gil-Marin, Anthony Gonzales, Rafael Guzman, Marc Kamionkowski, Alexie Leauthaud, Nikhil Padmanabhan, Uros Seljak, Ata Sarajedini, Vicki Sarajedini, Subir Sarkar, Kendrick Smith, Charles Telesco, Richard Woodard, and Wei Xue for wise words on the science and the overall planning of this analysis.

%Required DESI ack is below.
This material is based upon work supported by the U.S.\ Department of Energy (DOE), Office of Science, Office of High-Energy Physics, under Contract No.\ DE–AC02–05CH11231, and by the National Energy Research Scientific Computing Center, a DOE Office of Science User Facility under the same contract. Additional support for DESI was provided by the U.S. National Science Foundation (NSF), Division of Astronomical Sciences under Contract No.\ AST-0950945 to the NSF’s National Optical-Infrared Astronomy Research Laboratory; the Science and Technology Facilities Council of the United Kingdom; the Gordon and Betty Moore Foundation; the Heising-Simons Foundation; the French Alternative Energies and Atomic Energy Commission (CEA); the National Council of Humanities, Science and Technology of Mexico (CONAHCYT); the Ministry of Science, Innovation and Universities of Spain (MICIU/AEI/10.13039/501100011033), and by the DESI Member Institutions: \url{https://www.desi.lbl.gov/collaborating-institutions}. Any opinions, findings, and conclusions or recommendations expressed in this material are those of the author(s) and do not necessarily reflect the views of the U.S.\ National Science Foundation, the U.S.\ Department of Energy, or any of the listed funding agencies.

Research at the Perimeter Institute is supported in part by the Government of Canada through the Department of Innovation, Science and Economic Development Canada and by the Province of Ontario through the Ministry of Colleges and Universities.

The authors are honored to be permitted to conduct scientific research on I'oligam Du'ag (Kitt Peak), a mountain with particular significance to the Tohono O’odham Nation.

\bibliography{ref}

\end{document}